\documentclass[letterpaper]{article} 
\usepackage{aaai25}  
\usepackage{times}  
\usepackage{helvet}  
\usepackage{courier}  
\usepackage[hyphens]{url}  
\usepackage{graphicx} 
\usepackage{natbib}  
\usepackage{caption} 
\usepackage{algorithm}
\usepackage{algorithmic}

\usepackage{amsmath}

\usepackage{tabularx}
\usepackage{booktabs}
\usepackage{subcaption}
\usepackage{multirow}

\usepackage[T1]{fontenc}

\usepackage{xcolor}
\newcommand{\answerYes}[1]{\textcolor{blue}{#1}} 
 
\newcommand{\answerNA}[1]{\textcolor{gray}{#1}}

\usepackage{newfloat}
\usepackage{listings}
\DeclareCaptionStyle{ruled}{labelfont=normalfont,labelsep=colon,strut=off} 
\floatstyle{ruled}
\newfloat{listing}{tb}{lst}{}
\floatname{listing}{Listing}
\title{Understanding Online Polarization Through Human-Agent Interaction in a Synthetic LLM-Based Social Network}

\author {
    Tim Donkers\textsuperscript{\rm 1},
    J{\"u}rgen Ziegler\textsuperscript{\rm 1}
}
\affiliations {
    \textsuperscript{\rm 1}University of Duisburg-Essen\\
    tim.donkers@uni-due.de, juergen.ziegler@uni-due.de
}

\usepackage{bibentry}

\begin{document}

\maketitle

\begin{abstract}
The rise of social media has fundamentally transformed how people engage in public discourse and form opinions. While these platforms offer unprecedented opportunities for democratic engagement, they have been implicated in increasing social polarization and the formation of ideological echo chambers. Previous research has primarily relied on observational studies of social media data or theoretical modeling approaches, leaving a significant gap in our understanding of how individuals respond to and are influenced by polarized online environments. Here we present a novel experimental framework for investigating polarization dynamics that allows human users to interact with LLM-based artificial agents in a controlled social network simulation. Through a user study with 122 participants, we demonstrate that this approach can successfully reproduce key characteristics of polarized online discourse while enabling precise manipulation of environmental factors. Our results provide empirical validation of theoretical predictions about online polarization, showing that polarized environments significantly increase perceived emotionality and group identity salience while reducing expressed uncertainty. These findings extend previous observational and theoretical work by providing causal evidence for how specific features of online environments influence user perceptions and behaviors. More broadly, this research introduces a powerful new methodology for studying social media dynamics, offering researchers unprecedented control over experimental conditions while maintaining ecological validity.
\end{abstract}

\section{Introduction}
\label{sec:intro}

Online social networks have profoundly reshaped how individuals consume information and form opinions, offering both new possibilities for democratic engagement and novel challenges to societal cohesion. Although a substantial body of research has mapped the prevalence of social polarization---often characterized by the formation of ideologically distinct groups and limited cross-group interaction \cite{grover_dilemma_2022, kubin_role_2021, bail_exposure_2018}---much of this knowledge stems from two main methodological streams. Observational studies harness large-scale social media data through sentiment analysis \cite{karjus_evolving_2024, alsinet_measuring_2021, buder_does_2021}, network clustering \cite{treuillier_gaining_2024, bond_political_2022, al_amin_unveiling_2017}, or topic modeling \cite{kim_polarized_2019, chen_modeling_2021}. These approaches reveal macro-level patterns but cannot fully disentangle causal factors. In parallel, theoretical simulations and opinion dynamics models \cite{hegselmann_opinion_2002,degroot_reaching_1974,sasahara_social_2021,del_vicario_modeling_2017} allow for more controlled experimentation, yet tend to rely on simplified interaction rules that underrepresent the complexities of real human communication.

Meanwhile, advances in large language model (LLM) technology have opened new avenues for simulating online discourse and opinion formation \cite{chuang_simulating_2024, breum_persuasive_2024, ohagi_polarization_2024}. Despite this progress, comparatively little work has involved experimental studies with human participants interacting alongside these artificial agents. Such user studies could offer direct evidence of how polarized environments influence individual attitudes, social perceptions, and engagement patterns, thus illuminating mechanisms that purely observational or simplified simulated data cannot capture.

In this paper, we address this gap by proposing a novel experimental framework that integrates LLM-based agents and human participants within a synthetic social network platform. By focusing on message posting, interaction modeling, and recommendation mechanisms, we recreate a realistic microcosm of polarized discourse. We then collect detailed pre- and post-interaction survey data on participants' opinions, platform perceptions, and behavioral tendencies to map how exposure to polarized content and LLM-driven interactions shapes users’ subsequent viewpoints.

Our research makes the following key contributions:

\begin{enumerate}
    \item \textbf{Synthetic Network Testbed:} We develop and deploy a platform where LLM-based agents produce contextually relevant, polarized content, providing a controlled yet realistic environment for human experimentation.
    \item \textbf{Experimental Insights:} Through our user study, we track how polarized discussions influence emotional engagement, identity salience, and willingness to express uncertainty, offering empirical insights into mechanisms of online polarization.
    \item \textbf{Future Research Framework:} Our approach serves as a replicable template for investigating other facets of online social dynamics---including recommendation bias, echo chamber formation, and potential intervention strategies.
\end{enumerate}

Empirical findings from our study indicate that exposure to polarized exchanges significantly amplifies participants’ emotional responses and heightens perceived in-group identities. Moreover, recommendation bias was found to interact with polarized content to further skew user engagement, thus underscoring the need for experimental methods in the ongoing investigation of online polarization. 

The remainder of this paper is organized as follows. Section~\ref{sec:related-work} briefly reviews related research on online polarization and LLM-based simulations. Section~\ref{sec:simulation-model} introduces our simulation model and agent design. Section~\ref{sec:user-study} details the methodology, metrics, and outcomes of our user study. Section~\ref{sec:overall-discussion} discusses broader implications, and Section~\ref{sec:conclusion} concludes with suggestions for future work.

\section{Related Work}
\label{sec:related-work}

Research on social polarization in online networks spans multiple disciplines and methodologies, primarily falling into three streams: observational studies, theoretical modeling, and experimental user studies.

Observational studies leverage large-scale social media data to map polarization patterns. Early work identified ideologically segregated networks and the role of user choice and algorithms in echo chambers \citep{conover_political_2011, bakshy_exposure_2015}. Subsequent research confirmed these patterns across diverse platforms \citep{yarchi_political_2021} and explored user roles in information dissemination \citep{jiang_political_2020}. While revealing macro-level trends, these studies face challenges in disentangling underlying micro-mechanisms.

Theoretical models simulate opinion dynamics to understand how collective patterns emerge from individual interactions. Classical models, such as weighted averaging \citep{degroot_reaching_1974} and bounded confidence frameworks \citep{hegselmann_opinion_2002}, established basic mechanisms leading to consensus or opinion clustering but often relied on simplified interaction rules. Modern extensions incorporate greater psychological realism by modeling cognitive biases like confirmation bias \citep{del_vicario_modeling_2017}, network rewiring dynamics \citep{sasahara_social_2021}, and the crucial role of algorithmic amplification in shaping exposure and reinforcing divisions \citep{donkers_-sounding_2023, geschke_triple-filter_2019}. Some work in this area specifically explores recommendation strategies aimed at mitigating polarization or fostering diversity \citep{stray_designing_2022,donkers_dual_2021, garimella_reducing_2017}. Overall, these computational opinion dynamics models offer methodologically rigorous insights into specific mechanisms but typically abstract away the complexities of natural language communication that characterize real-world online discourse.

Recent advances in large language models (LLMs) have spurred a new wave of simulation approaches. Integrating LLMs into agent-based modeling (ABM) allows for simulations with richer, language-based interactions, potentially enhancing the behavioral realism compared to traditional rule-based ABM \citep{ghaffarzadegan_generative_2024, gao_large_2024}. Studies have shown that LLM agents can replicate polarization dynamics \citep{ohagi_polarization_2024}, information diffusion, and emotional contagion through text exchanges \citep{gao_social_2024}, and respond to persuasive rhetorical strategies \citep{breum_persuasive_2024}. However, these agent-agent simulations often lack the direct theoretical grounding of classical models or rigorous validation with human behavior.

Experimental studies involving human participants offer unique potential for causal inference but remain relatively scarce compared to observational and modeling work. Building on foundations in social psychology regarding conformity and group identity \citep{asch_effects_1951, tajfel_social_1971}, contemporary experiments often focus on targeted interventions, examining how specific factors like content exposure \citep{banks_polarizedfeeds_2021} or social identity cues \citep{wuestenenk_influence_2023} affect perceptions or attitudes in controlled settings. Other studies use simplified experimental tasks (e.g., numeric estimations) to calibrate parameters for formal opinion dynamics models \citep{chacoma_opinion_2015, das_modeling_2014}. While valuable, these approaches typically either lack the ecological validity of a full social media environment or constrain interaction in ways that miss the nuances of online discourse.

Our work bridges these gaps by developing an experimental framework that integrates LLM-based agents generating realistic discourse within a controlled simulation environment, allowing human participants to interact naturally. This approach enables the systematic manipulation of environmental factors (like discourse polarization and recommendation bias) grounded in modeling principles, while measuring behavioral and perceptual responses in an ecologically relevant context, thus combining the strengths of controlled experimentation and realistic interaction.
\section{Simulation Model}
\label{sec:simulation-model}

Our computational framework integrates opinion dynamics principles with LLM-based agents to investigate polarization in social networks. The framework employs OpenAI's GPT-4o-mini model\footnote{\url{https://platform.openai.com/docs/overview}} to create agents capable of generating realistic content through posts, comments, and reposts, as well as performing social media interactions such as liking and following. The system architecture comprises three primary components: an agent-based architecture, an LLM infrastructure for content generation, and a social network structure for information dissemination.

Each agent in our simulation represents an individual user, characterized by a continuous opinion value $o_i \in [-1,1]$ (negative=opposition, positive=support), a personality description, a biography, and an interaction history. The initial opinion distribution is configurable: for polarized conditions, a bimodal distribution (e.g., centered at $\pm 0.8$) is used, while for moderate conditions, a unimodal distribution (e.g., centered at 0) is employed (see Appendix \ref{app:simulation-model} for details). Agents are also distinguished as regular users or influencers, with influencers having higher posting probabilities ($p_{\text{inf}} > p_{\text{reg}}$) and greater network influence.

Message generation relies on LLM prompting structured by multiple contextual elements: the agent's specific opinion value ($o_i$), the topic description, their personality, recent interactions, and opinion intensity-based instructions (e.g., expressing more certainty and emotion for higher $|o_i|$). This comprehensive context ensures messages reflect individual characteristics and appropriate discourse patterns. Detailed prompting components are listed in Appendix \ref{app:simulation-model}, Table \ref{tab:prompt-components}.

Agent interactions (likes, comments, reposts) are governed by a probabilistic model, $P_{\text{react}}(A_i, m)$, determining the likelihood of agent $A_i$ reacting to a message $m$. The interaction probability is formalized as:

\begin{align}
\label{eq:reaction-prob}
P_{\text{react}}(A_i, m) =\; &p_b \cdot \big((1-w) + w \cdot \psi(o_i)\big) \cdot \nonumber \\
&\rho(o_i, \pi(A_i, m))
\end{align}

This equation combines three key components: (1) a base probability ($p_b$) specific to the interaction type (like, comment, repost); (2) the agent's opinion strength ($\psi(|o_i|)$), where agents with stronger opinions have a higher baseline reactivity, weighted by parameter $w$; and (3) an opinion alignment function $\rho(o_i, \pi(A_i, m))$.

The function $\pi(A_i, m)$ uses an LLM to assess the opinion expressed in message $m$ relative to agent $A_i$'s context, mapping it to the $[-1,1]$ scale. The alignment function $\rho$ (detailed in Appendix \ref{app:simulation-model}) calculates the reaction likelihood based on the similarity ($\rho_{\text{sim}}$) or difference ($\rho_{\text{opp}}$) between the agent's opinion $o_i$ and the assessed message opinion $\pi(A_i, m)$. It uses sigmoid functions to model a continuous transition, resulting in a U-shaped probability curve (see Appendix \ref{app:simulation-model}, Figure \ref{fig:reaction-function}): interactions are likely for high agreement ($\rho_{\text{sim}}$ dominates) and, to a controlled extent, for high disagreement ($\rho_{\text{opp}}$ dominates), but unlikely for neutral alignment. Crucially, the influence of disagreement is modulated by a cross-ideology factor $c \in [0,1]$, which is set higher for comments and reposts ($c>0$) than for likes ($c=0$), enabling cross-ideological engagement for more active interaction types.

The social network is modeled as a directed graph where nodes are agents and edges represent 'follow' relationships. The network structure evolves dynamically. Initial connections are formed probabilistically based on opinion homophily (agents are more likely to follow others with similar opinions) and influencer status (influencers are more likely to be followed). Subsequently, agents can form new connections based on recommendations and dissolve existing ones independently. Following decisions use the same reaction probability function (Eq.~\ref{eq:reaction-prob}) that governs other interactions, but with the agent's opinion values directly substituted for the message evaluation: $\rho(o_i, o_j)$ instead of $\rho(o_i, \pi(A_i, m))$. Unfollowing occurs randomly with a fixed base probability.

Information propagation is governed by a recommendation system determining message visibility. An influence score $\eta(A_j)$ is calculated for each potential message author $A_j$ based on their normalized follower count: $\eta(A_j) = |\text{followers}(A_j)| / (|\mathcal{V}| - 1)$, where $\mathcal{V}$ is the set of all agents. For a target agent $A_i$, the system recommends the top $N$ messages from the set of eligible messages (those not yet seen by $A_i$ and not authored by $A_i$) authored by agents with the highest influence scores $\eta$. This maintains controlled exposure while incorporating network structure effects.

The simulation operates through discrete time steps, with each iteration encompassing content generation, agent interactions (based on the probabilistic model and recommendations), and network evolution. This allows for examining emergent phenomena like network polarization and echo chamber formation. Further mathematical formalizations, specific parameter settings, and the simulation algorithm pseudocode are provided in Appendix \ref{app:simulation-model}.
\section{User Study}
\label{sec:user-study}

In this section, we will investigate how discussion polarization and algorithmic content curation in social media environments affect human perception of debates and their engagement behavior. Through a controlled experiment using a simulated social media platform, we examine how different levels of discussion polarization (polarized vs. moderate) and recommendation bias (pro, balanced, contra) shape opinion formation and interaction patterns.

\subsection{Experimental Design}

We employed a $2 \times 3$ between-subjects factorial design to investigate the dynamics of opinion formation and perception of polarization in online discussions. The experimental design manipulated two key dimensions: the \emph{Polarization Degree} in the artificial agent population and a systematic \emph{Recommendation Bias} while maintaining Universal Basic Income (UBI) as the consistent discussion topic. This design allows us to compare the effects of high polarization against a baseline of moderate polarization. Our focus is specifically on understanding how increasing the intensity of polarization in the discourse environment affects user perception and behavior.

The first experimental dimension contrasted highly polarized discussions with a moderate discourse condition through the manipulation of artificial agent behavior. In the polarized condition, artificial agents expressed extreme viewpoints and employed confrontational discourse patterns, characterized by emotional language, strong assertions, and minimal acknowledgment of opposing viewpoints. The moderate condition, serving as the comparative baseline, featured more nuanced discussions and cooperative interaction styles, with agents expressing uncertainty, acknowledging limitations in their knowledge, and engaging constructively with opposing views.

The second dimension introduced systematic bias in the recommendation system, implemented across three levels: neutral ($50\%$ pro-UBI, $50\%$ contra-UBI content), pro-bias ($70\%$ pro-UBI, $30\%$ contra-UBI content), and contra-bias ($30\%$ pro-UBI, $70\%$ contra-UBI content). This manipulation aimed to investigate how algorithmic content curation interacts with the discourse environment to potentially shape opinion formation and perception of debate polarization.

The selection of UBI as the focal topic was driven by several strategic considerations. Unlike heavily polarized topics where individuals often hold entrenched positions, UBI represents an emerging policy proposal where public opinion remains relatively malleable, making it ideal for studying opinion formation and polarization dynamics. While UBI evokes fewer preset opinions, it remains sufficiently concrete and consequential to generate meaningful discourse, with its complexity spanning economic, social, and technological dimensions. Recent polling data supports UBI's suitability, showing a balanced distribution of opinions with approximately $51.2\%$ of Europeans \cite{vlandas_politics_2019} and $48\%$ of Americans \cite{hamilton_people_2022} expressing support, while significant portions remain undecided or hold moderate views. Additionally, UBI's limited real-world implementation means participants' opinions are more likely to be based on theoretical arguments rather than direct experience or partisan allegiances, allowing us to examine how social media dynamics can influence opinion formation before entrenched polarization takes hold.

\subsection{System Prototype}

The prototype implementation consists of a web application that simulates a social media platform, reminiscent of X (formerly Twitter), to study social polarization dynamics. The interface, as depicted in Appendix \ref{app:user-study}, Figure~\ref{fig:prototype-screenshot}, adheres to a familiar social media layout, facilitating user engagement and interaction.

The application's main interface is divided into three primary sections: a navigation sidebar on the left, a central Newsfeed, and a recommendation panel on the right. The navigation sidebar provides quick access to essential functionalities such as the user's profile, a general user overview, and a logout option. The central Newsfeed serves as the primary interaction space, where users can view and engage with recommended posts from other users (see the appendix for details on the recommendation procedure). At the top of the Newsfeed, a text input area invites users to share their thoughts, mimicking the spontaneous nature of social media communication.

The Newsfeed displays a series of posts, each accompanied by user avatars, usernames, timestamps, and interaction metrics such as \emph{likes}, \emph{comments}, and \emph{reposts}. This design encourages user engagement and provides visual cues about the popularity and impact of each post. The recommendation panel on the right side of the interface suggests other users to follow, potentially influencing the user's network expansion and exposure to diverse viewpoints.

User profiles are dynamically generated, displaying the user's posts, follower relationships, and other relevant metadata like a user's handle and biography. It is also possible to follow and unfollow artificial users.

\subsection{Procedure}
The experiment consisted of three phases: pre-interaction, interaction, and post-interaction.

\paragraph{Pre-interaction Phase} Participants first completed a comprehensive questionnaire assessing various baseline measures. These included demographic information, social media usage patterns, and initial attitudes towards UBI. 

\paragraph{Interaction Phase} Afterwards, participants were introduced to our simulated social media platform. Each participant interacted individually in an isolated instance of the platform, i.e., they did not communicate with other human users. Instead, the platform was populated by a set of 30 pre-programmed artificial agents specific to their assigned experimental condition, including three designated 'influencers' on each side of the UBI debate. These agents first engaged in agent-only simulations run for $10$ iterations. These preparatory simulations, conducted under either polarized or moderate initial opinion distributions with agent opinions held fixed, generated the platform's initial state, including a history of posts, comments, reposts, likes, and follow relationships. 

In the polarized condition, agents averaged $2.16$ posts, $8.07$ likes, $5.47$ comments, and $2.90$ reposts per agent, with influencers showing notably higher activity. The moderate condition showed similar posting patterns ($2.07$ posts per agent) but reduced commenting ($1.60$) and reposting ($1.57$) activity. Full statistics are available in Appendix B, Table~\ref{tab:agent_platform_stats}.

Participants were instructed to engage with the platform naturally, as they would in their regular social media use. They received the following instructions:
\begin{quote}
"You will now interact with a social media platform discussing \emph{Universal Basic Income}. Please use the platform as you normally would use social media. You can read posts, like them, comment on them, or create your own posts. Your goal is to form an opinion on the topic. You will have 10 minutes for this task."
\end{quote}

During this 10-minute phase, participants' interactions (likes, comments, reposts, follows) were recorded for later analysis. Crucially, while participants interacted, the artificial agents did not generate any new content; this was a deliberate design choice to ensure exposure to a controlled initial environment. However, the participant's newsfeed remained dynamic, updated by the recommendation algorithm which presented content from the pre-generated pool according their individual interaction behavior.

\paragraph{Post-interaction Phase} After the interaction period, participants completed a post-test questionnaire. This included measures of their perception of the key constructs listed below. Additionally, participants evaluated the realism and effectiveness of the simulated platform.

\subsubsection{Ethical Considerations and Risk Mitigation}

While our primary goal is to better understand and mitigate harmful polarization, we recognize that such simulations could potentially cause harm to study participants. In accordance with the review of our institution's ethics committee, we have taken several precautionary steps to address potential risks:

\begin{itemize}
    \item Data Anonymization and Participant Welfare: All participant data were anonymized, and recruitment processes adhered to established privacy and consent guidelines. Participants were fully debriefed about the nature of the artificial agents and could withdraw from the study at any time without penalty.

    \item Responsible Platform Design: While we aimed to mimic real social media interfaces to enhance ecological validity, we integrated disclaimers in participant instructions, clarifying the experimental setting. This reduces deception and protects against unintended distress.
\end{itemize}

\subsection{Measures}

All constructs were measured using four-item scales rated on 5-point Likert scales (1 = Strongly Disagree to 5 = Strongly Agree) (see Appendix \ref{app:user-study}, Table~\ref{tab:factor-loadings} for the full item list and factor loadings). Key constructs measured in this study included:

\begin{itemize}
    \item \emph{Opinion Change}: Measured shifts in participants' opinions about UBI between pre- and post-interaction phases. This was calculated as the difference between the average scores of the opinion items before and after interaction.
    
    \item \emph{Perceived Polarization}: Assessed participants' perception of opinion extremity and ideological division, focusing on the perceived distance between opposing viewpoints.
    
    \item \emph{Perceived Group Salience}: Evaluated the extent to which participants perceived the discussion as being driven by group identities rather than individual perspectives. Unlike other measures, this scale retained only two of four initial items after psychometric analysis (see Appendix \ref{app:user-study}).
    
    \item \emph{Perceived Emotionality}: Measured participants' assessment of the emotional intensity and affective tone, capturing the perceived level of emotional versus rational discourse.
    
    \item \emph{Perceived Uncertainty}: Captured the degree to which participants observed expressions of doubt and acknowledgment of knowledge limitations.
    
    \item \emph{Perceived Bias}: Evaluated participants' assessment of viewpoint balance and fair representation.
\end{itemize}

\subsection{Participants}

We recruited $122$ participants through the Prolific\footnote{https://www.prolific.com/} platform (compensation was set according the recommended pay rate of $\$12$/hour). Participants were evenly distributed across the six experimental conditions (N $\approx$ 20 per condition). The sample exhibited a gender distribution favoring male participants ($63.6\%$) over female participants ($35.7\%$), with a single participant preferring not to disclose their gender. The age distribution revealed a predominantly young to middle-aged sample, with approximately $61.5\%$ of participants falling between $20$ and $39$ years old. The modal age group was $25$-$29$ years (18.6\%), followed by $35$-$39$ years ($15.0\%$).

Regarding educational background, nearly half of the participants ($49.3\%$) held university degrees, indicating a relatively high level of formal education in the sample. The remaining participants were distributed across various educational qualifications, with A-levels/IB, GCSE, and vocational certifications each representing approximately $11\%$ of the sample.

The majority of participants were professionally active, with $58.6\%$ being employees and $10.7\%$ self-employed. The sample also included a notable proportion of students ($15.7\%$ combined university and school students), reflecting diverse occupational backgrounds.

Participants demonstrated high engagement with social media platforms, with $80\%$ reporting daily or near-constant usage. The majority ($75\%$) spent between one and four hours daily on social media platforms. YouTube ($25.0\%$) and Facebook ($24.3\%$) emerged as the most frequently used platforms, followed by Instagram ($17.9\%$) and X, formerly Twitter ($10.7\%$). This usage pattern suggests participants were well-acquainted with social media interfaces and interaction patterns, making them suitable subjects for the study's simulated social media environment.

\subsection{Analysis of Debate Perception}

Our first analysis examines how users perceive and process discussions under varying conditions of \emph{Polarization Degree} and \emph{Recommendation Bias} (see Appendix \ref{app:user-study} for preliminary analyses). We specifically investigate whether participants recognize polarized discourse patterns, how they process emotional and group-based content, and how \emph{Recommendation Bias} might moderate these perceptions. Through this analysis, we aim to understand the psychological mechanisms through which discussion climate and content curation are associated with users' experience of online debates.

\begin{table*}[ht]
\footnotesize
\centering
\begin{tabularx}{\textwidth}{>{\raggedright\arraybackslash}p{2.5cm}>{\raggedright\arraybackslash}p{1.5cm}*{3}{>{\centering\arraybackslash}X}>{\centering\arraybackslash}p{1.2cm}>{\centering\arraybackslash}p{1.2cm}>{\centering\arraybackslash}p{1.2cm}}
\toprule
\multirow{2}{*}{\textbf{Metric}} & \multirow{2}{*}{\textbf{Polarization}} & \multicolumn{3}{c}{\textbf{Recommendation Bias}} & \multicolumn{3}{c}{\textbf{ANOVA $\boldsymbol{F}$($\boldsymbol{p}$)}} \\
\cmidrule(lr){3-5} \cmidrule(lr){6-8}
& & Contra & Balanced & Pro & Pol & Rec & Pol×Rec \\
\midrule
\multirow{2}{*}{Opinion Change} 
   & Polarized & \textbf{-0.41} (0.60) & -0.08 (0.55) & -0.10 (0.33) & 3.48 & 1.74 & 1.90 \\
   & Moderate & -0.06 (0.38) & -0.09 (0.43) & -0.02 (0.26) & (.065) & (.180) & (.154) \\
\midrule
\multirow{2}{*}{Polarization} 
   & Polarized & \textbf{4.47} (0.82) & \textbf{4.61} (0.97) & \textbf{4.29} (0.75) & \textbf{56.48***} & 0.02 & 1.50 \\
   & Moderate & 3.30 (0.69) & 3.23 (0.83) & 3.54 (0.84) & (.000) & (.979) & (.228) \\
\midrule
\multirow{2}{*}{Emotionality} 
   & Polarized & \textbf{3.80} (0.48) & \textbf{3.59} (0.99) & \textbf{3.36} (0.80) & \textbf{73.54***} & 1.42 & 2.44 \\
   & Moderate & 2.50 (0.74) & 2.17 (0.60) & 2.65 (0.83) & (.000) & (.245) & (.092) \\
\midrule
\multirow{2}{*}{Group Salience} 
   & Polarized & \textbf{2.94} (0.60) & \textbf{2.86} (0.51) & \textbf{3.21} (0.42) & \textbf{17.18***} & \textbf{5.63**} & 0.07 \\
   & Moderate & 2.51 (0.59) & 2.42 (0.65) & 2.86 (0.40) & (.000) & (.005) & (.934) \\
\midrule
\multirow{2}{*}{Bias} 
   & Polarized & \textbf{3.50} (0.87) & 3.20 (0.76) & \textbf{3.83} (0.59) & \textbf{15.08***} & 0.44 & \textbf{3.73*} \\
   & Moderate & 3.05 (0.82) & 3.02 (0.87) & 2.68 (0.87) & (.000) & (.648) & (.027) \\
\midrule
\multirow{2}{*}{Uncertainty} 
   & Polarized & 2.29 (0.60) & 2.03 (0.60) & 1.90 (0.75) & \textbf{48.86***} & 0.97 & 1.06 \\
   & Moderate & \textbf{2.92} (0.71) & \textbf{2.83} (0.69) & \textbf{2.96} (0.48) & (.000) & (.383) & (.350) \\
\bottomrule
\multicolumn{8}{p{.95\textwidth}}{\small \textbf{Note:} Values show means with standard deviations in parentheses. Bold values indicate significantly higher means between polarized and Moderate conditions. F-statistics in bold are significant. Pol = Polarization main effect, Rec = Recommendation main effect, Pol×Rec = Interaction effect. *$p$ < .05, **$p$ < .01, ***$p$ < .001} \\
\end{tabularx}
\caption{Effects of Polarization and Recommendation Bias on Key Dependent Variables}
\label{tab:perception-anova}
\end{table*}

\subsubsection{Results}
\begin{figure*}[h]
    \centering
    \includegraphics[width=\textwidth]{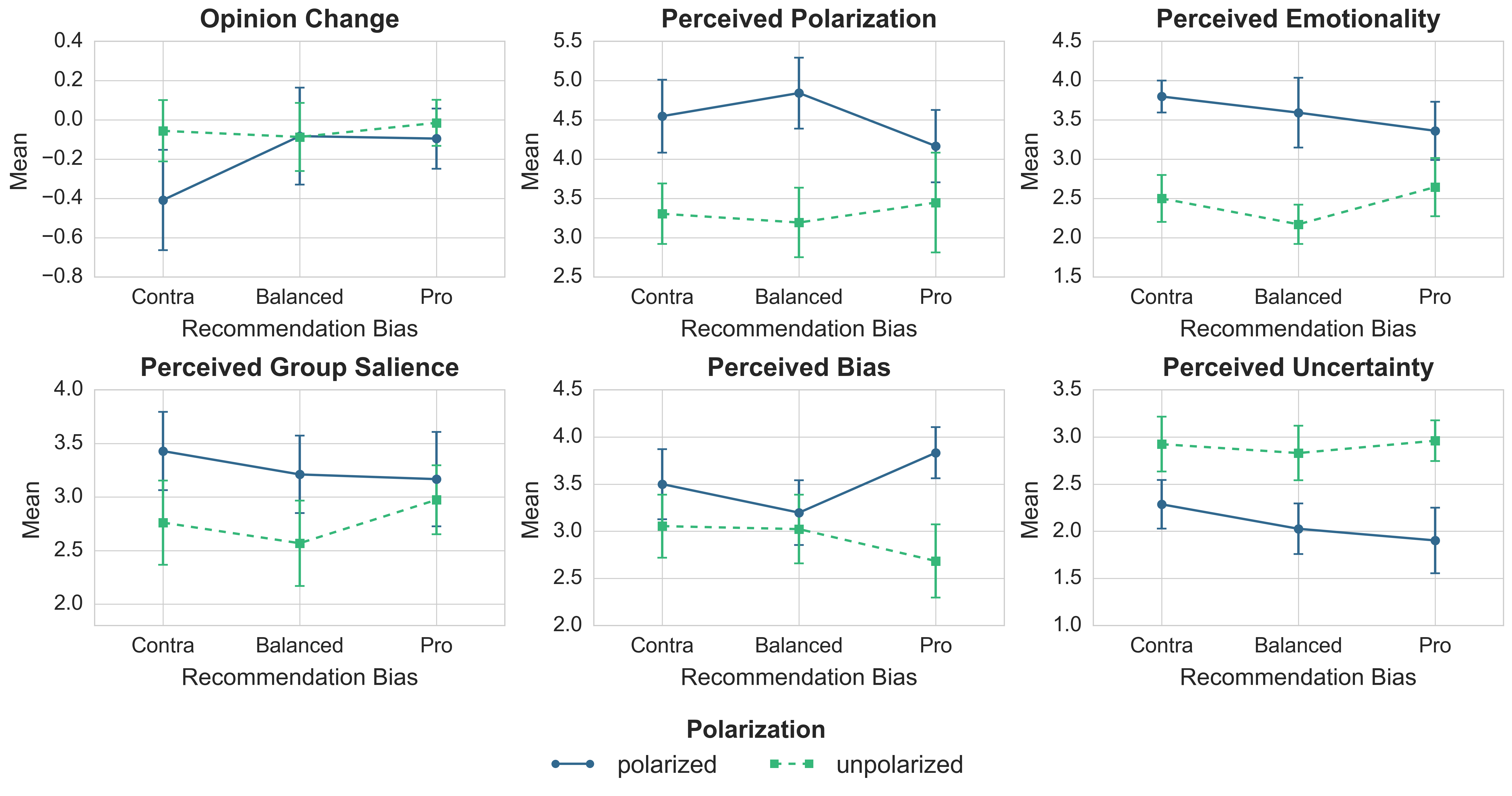}
    \caption{Interaction plots showing the effects of polarization and recommendation type on dependent variables. Blue lines represent the polarized condition, green lines represent the unpolarized condition. Error bars represent $95\%$ confidence intervals.}
    \label{fig:perception-interaction-plots}
\end{figure*}

We analyzed the effects of discussion \emph{Polarization Degree} and \emph{Recommendation Bias} using $2 \times 3$ ANOVAs (see Table~\ref{tab:perception-anova} for full statistics and Figure~\ref{fig:perception-interaction-plots} for interaction plots). The results consistently showed significant main effects of \emph{Polarization Degree} across most perceptual variables, whereas \emph{Recommendation Bias} had a more limited impact.

Regarding \emph{Opinion Change}, neither \emph{Polarization Degree} nor \emph{Recommendation Bias} yielded statistically significant main effects (Table~\ref{tab:perception-anova}). However, descriptive statistics suggested a trend towards stronger opinion shifts (more negative) specifically in the polarized condition combined with contra-bias recommendations ($M = -0.41$), compared to minimal changes in other conditions.

The \emph{Polarization Degree} manipulation strongly influenced the perceived discussion climate. Participants rated \emph{Perceived Polarization} significantly higher in the polarized condition ($M = 4.46$) compared to the unpolarized condition ($M = 3.35$), a difference confirmed by post-hoc tests ($p < .001$, Hedges' $g = 1.36$). This represented a substantial main effect (Table~\ref{tab:perception-anova}).

\emph{Perceived Emotionality} showed the strongest main effect of \emph{Polarization Degree} in the study ($\eta_p^2 = 0.388$, $p < .001$), with polarized discussions perceived as significantly more emotional ($M = 3.58$ vs. $M = 2.44$ for unpolarized; Hedges' $g = 1.53$). While the main effect of \emph{Recommendation Bias} was not significant, a marginal interaction effect with \emph{Polarization Degree} was observed ($p = .092$).

Similarly, \emph{Perceived Uncertainty} was significantly lower in polarized discussions ($M = 2.07$) compared to unpolarized ones ($M = 2.90$), indicating another strong main effect of \emph{Polarization Degree} ($\eta_p^2 = 0.296$, $p < .001$; Hedges' $g = -1.26$).

\emph{Perceived Group Salience} was significantly affected by both \emph{Polarization Degree} ($p < .001$) and \emph{Recommendation Bias} ($p = .005$). Polarized conditions elicited higher perceptions of group-based discourse ($M = 3.00$ vs. $M = 2.59$ for unpolarized; Hedges' $g = 0.73$). Post-hoc tests for \emph{Recommendation Bias} indicated significantly higher salience in the pro-bias condition compared to contra-bias ($p=.011$) and neutral ($p=.001$) conditions (see Table~\ref{tab:perception-anova} for means).

Finally, \emph{Perceived Bias} showed a significant main effect of \emph{Polarization Degree} ($p < .001$) and a significant interaction effect ($p = .027$). Participants perceived higher bias in the polarized condition ($M = 3.51$) compared to the unpolarized condition ($M = 2.92$; Hedges' $g = 0.69$). The interaction indicated this effect was particularly pronounced under pro-bias recommendations (Table~\ref{tab:perception-anova}).

\subsubsection{Discussion} 

Our findings reveal significant insights into how users perceives and process polarized discussions within our simulated social media environment. The results demonstrate that the manipulation of \emph{Polarization Degree} was associated with profound differences across multiple perceptual dimensions, while \emph{Recommendation Bias} played a more nuanced role in shaping user experiences in this context.

The strong association between \emph{Polarization Degree} and \emph{Perceived Emotionality} ($\eta_p^2 = 0.388$) suggests that participants were highly attuned to the emotional tenor of discussions. This heightened emotional perception in polarized conditions indicates that users readily distinguish between measured discourse and emotionally charged exchanges, which might relate to their willingness to engage in discussions. The parallel increase in \emph{Perceived Polarization} ($\eta_p^2 = 0.327$) further supports this interpretation, suggesting that emotional intensity may serve as a key signal for users in assessing the degree of polarization in online discussions they encounter.

A particularly noteworthy finding is the inverse relationship between \emph{Polarization Degree} and \emph{Perceived Uncertainty} ($\eta_p^2 = 0.296$). The significant decrease in \emph{Perceived Uncertainty} within polarized discussions aligns with theoretical frameworks suggesting that polarized discourse often manifests through increased assertiveness and reduced acknowledgment of epistemic limitations. This pattern observed in our experiment may help explain the self-reinforcing nature often attributed to polarized discussions: as uncertainty expressions diminish, the space for nuanced dialogue and opinion adjustment potentially contracts.

The observed differences on \emph{Perceived Group Salience} ($\eta_p^2 = 0.129$) and \emph{Perceived Bias} ($\eta_p^2 = 0.115$) provide insights into how polarization might relate to the social-cognitive dimensions of online discourse. The emergence of stronger group-based discourse perceptions in polarized conditions suggests that heightened polarization could be linked to increased social categorization processes, potentially facilitating the formation of opposing camps. This interpretation is strengthened by the significant interaction between \emph{Polarization Degree} and \emph{Recommendation Bias} on \emph{Perceived Bias}, particularly pronounced in pro-bias conditions. This interaction suggests that algorithmic curation, when combined with polarized discourse in this manner, may amplify the salience of group divisions.

The absence of significant effects on \emph{Opinion Change}, despite the substantial perceptual differences across conditions, merits careful consideration. This finding suggests that while users readily recognize polarized discourse patterns, their own positions might remain relatively stable during the kind of short-term exposure used in this study. However, the trending pattern of stronger opinion changes in polarized conditions with contra-bias hints at potential longer-term effects that warrant further investigation.

These results carry important implications for thinking about platform design and intervention strategies. The clear user sensitivity to emotional content and group-based discourse suggests that interventions targeting these aspects might be particularly effective in mitigating polarization effects. Moreover, the complex interaction between \emph{Polarization Degree} and \emph{Recommendation Bias} indicates that algorithmic content curation strategies should consider not only viewpoint diversity but also the emotional and social dynamics of the content being recommended.

Our findings also highlight the multifaceted nature of polarization perception. The strong correlations between emotional intensity, reduced uncertainty, and increased group salience suggest that polarization manifests through a constellation of interrelated perceptual changes. This insight suggests that effective interventions might need to address multiple aspects simultaneously rather than focusing on single dimensions of polarized discourse. It is important to reiterate, however, that these findings stem from a controlled experiment, and their direct applicability to real-world platform dynamics requires further investigation.

\begin{table*}[ht]
\centering
\small
\begin{tabularx}{\textwidth}{>{\raggedright\arraybackslash}p{2cm}>{\raggedright\arraybackslash}p{1.8cm}*{3}{>{\centering\arraybackslash}X}>{\centering\arraybackslash}p{1.2cm}>{\centering\arraybackslash}p{1cm}}
\toprule
\multirow{2}{*}{\textbf{Metric}} & \multirow{2}{*}{\textbf{Polarization}} & \multicolumn{3}{c}{\textbf{Recommendation Bias}} & \multicolumn{2}{c}{\textbf{ANOVA}} \\
\cmidrule(lr){3-5} \cmidrule(lr){6-7}
& & Contra & Balanced & Pro & $F$ & $p$ \\
\midrule
\multirow{2}{*}{Interactions} 
   & Polarized & \textbf{10.30} (12.43) & 4.82 (4.49) & 4.80 (5.29) & \multirow{2}{*}{\textbf{3.25}$^a$} & \multirow{2}{*}{.042*} \\
   & Moderate & 7.92 (7.62) & 4.71 (3.26) & \textbf{9.25} (12.85) & & \\
\midrule
\multirow{2}{*}{Likes} 
   & Polarized & \textbf{5.65} (9.53) & 2.55 (2.65) & 2.35 (2.54) & \multirow{2}{*}{2.25$^a$} & \multirow{2}{*}{.110} \\
   & Moderate & 4.38 (4.03) & 2.64 (2.44) & \textbf{4.95} (8.22) & & \\
\midrule
\multirow{2}{*}{Reposts} 
   & Polarized & 0.65 (1.03) & 0.45 (0.67) & 0.40 (0.82) & \multirow{2}{*}{1.89$^a$} & \multirow{2}{*}{.156} \\
   & Moderate & \textbf{1.08} (2.30) & 0.29 (0.46) & \textbf{1.10} (1.86) & & \\
\midrule
\multirow{2}{*}{Comments} 
   & Polarized & 0.35 (0.57) & 0.23 (0.43) & 0.25 (0.55) & \multirow{2}{*}{\textbf{6.51}$^b$} & \multirow{2}{*}{.012*} \\
   & Moderate & 0.50 (1.18) & 0.71 (1.12) & \textbf{1.10} (1.94) & & \\
\midrule
\multirow{2}{*}{Follows} 
   & Polarized & \textbf{3.30} (4.12) & 1.45 (1.84) & 1.50 (2.82) & \multirow{2}{*}{2.60$^a$} & \multirow{2}{*}{.078} \\
   & Moderate & 1.88 (3.38) & 1.04 (1.57) & 1.95 (3.33) & & \\
\bottomrule
\multicolumn{7}{p{.95\textwidth}}{\small \textbf{Note:} Values show means with standard deviations in parentheses. Bold values indicate highest means within recommendation bias conditions. $^a$F-statistic for main effect of Recommendation (df = 2, 134). $^b$F-statistic for main effect of Polarization (df = 1, 135). *$p$ < .05} \\
\end{tabularx}
\caption{Main Effects of Experimental Conditions on User Engagement}
\label{tab:interactions-anova}
\end{table*}

\begin{figure*}[h]
    \centering
    \includegraphics[width=\textwidth]{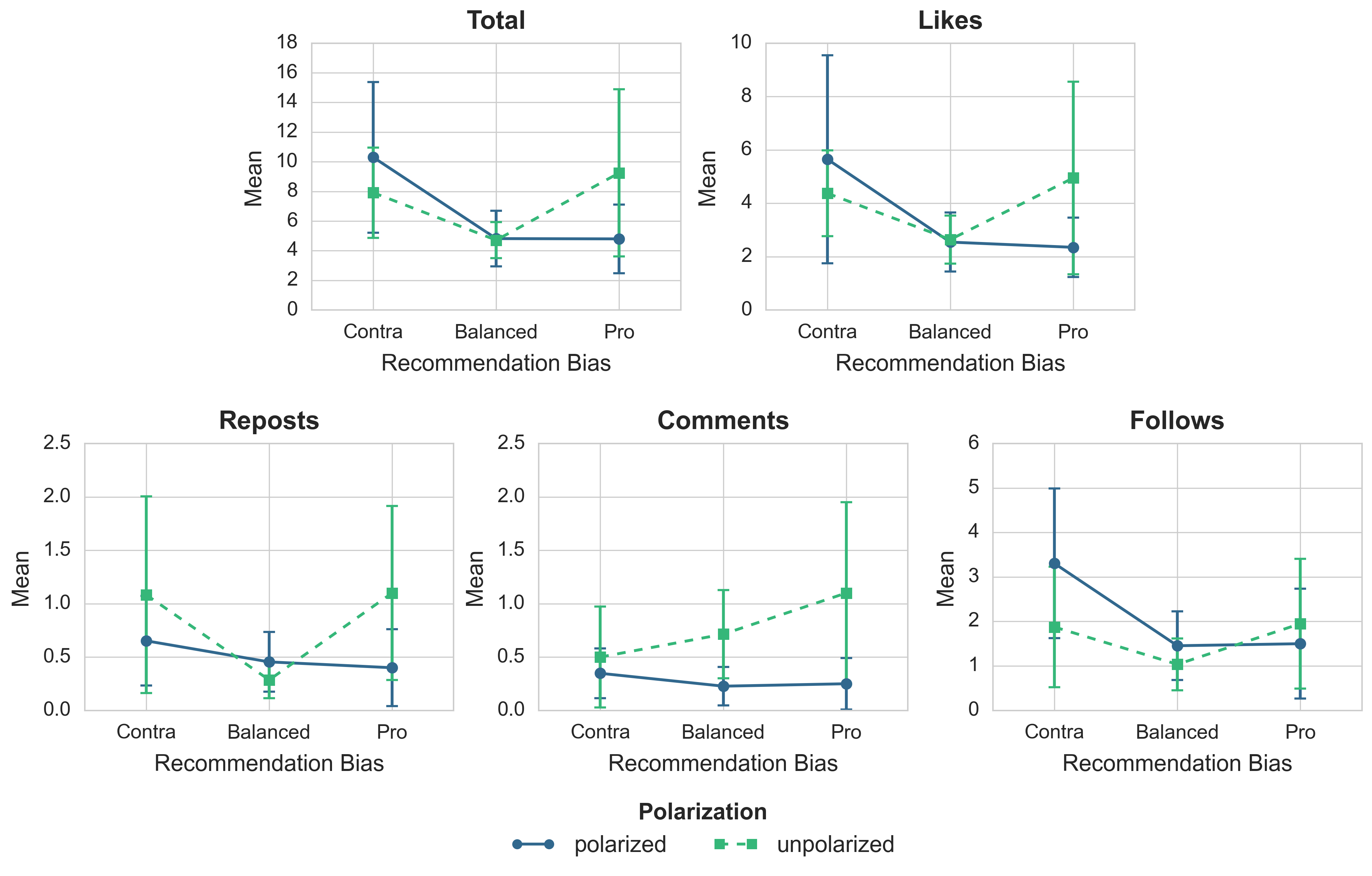}
    \caption{Interaction plots showing the effects of polarization and recommendation type on user engagement. Blue lines represent the polarized condition, green lines represent the unpolarized condition. Error bars represent $95\%$ confidence intervals.}
    \label{fig:interaction-plot-recommendation-bias}
\end{figure*}

\subsection{Analysis of User Engagement}

Our second analysis examines how polarization and recommendation bias shape user engagement behaviors on the platform. We investigate both the quantity and quality of interactions, analyzing how different experimental conditions affect users' preferences for specific types of engagement (\emph{likes}, \emph{comments}, \emph{reposts}, and \emph{follows}).  This analysis aims to understand whether polarized environments and algorithmic bias are associated with differences in not just how much users engage, but also how they choose to participate in discussions.

\subsubsection{Results}

Our analysis revealed distinct patterns of user engagement across the experimental conditions. Participants generated 946 interactions in total ($M = 6.91$, $SD = 8.50$, $Mdn = 5.00$, Range: $0-56$), with substantial individual variation. The majority ($81.02\%$) engaged at least once. Among active users, engagement levels spanned low ($1-5$ interactions; $36.5\%$), moderate ($6-10$; $24.82\%$), and high ($>10$; $19.71\%$) categories.

Figure~\ref{fig:stacked-interaction-distribution} shows a clear hierarchy in interaction types: \emph{Likes} were dominant ($54.02\%$, $M = 3.73$), followed by \emph{follows} ($26.53\%$, $M = 1.83$), \emph{reposts} ($9.41\%$, $M = 0.65$), and \emph{comments} ($7.61\%$, $M = 0.53$). This suggests a preference for lower-effort engagement.

Correlation analysis revealed strong positive associations between \emph{total interactions} and both \emph{likes} ($r = .93$, $p < .001$) and \emph{follows} ($r = .73$, $p < .001$), moderate correlation with \emph{comments} ($r = .45$, $p < .001$), and weaker correlation with \emph{reposts} ($r = .37$, $p < .001$). Notably, \emph{comments} and \emph{reposts} showed minimal correlation ($r = .01$, $p = .899$), suggesting distinct user purposes.

ANOVA results (Table~\ref{tab:interactions-anova}) indicated a significant main effect of \emph{Recommendation Bias} on \emph{total interactions} ($\eta^2_p = .046$). Post-hoc tests revealed significantly more interactions in the contra-bias condition compared to the balanced condition ($p = .008$, Hedges' $g = 0.56$); other comparisons were not significant (see Table~\ref{tab:interactions-anova} for means).

Analysis of specific interaction types showed a significant main effect of \emph{Polarization Degree} on \emph{commenting} ($\eta^2_p = .046$). Participants commented more in unpolarized conditions ($M = 0.77$) compared to polarized ones ($M = 0.28$), suggesting moderate environments may facilitate this form of engagement.

Other interaction types showed no significant main effects, though trends emerged. \emph{Following} behavior showed a marginal effect of \emph{Recommendation Bias} ($p = .078$, $\eta^2_p = .037$), with a tendency to follow more users when exposed to opposing views (contra-bias) compared to balanced content. \emph{Likes} showed a non-significant trend consistent with \emph{total interactions} ($p = .110$).

Examining the interaction patterns (Table~\ref{tab:interactions-anova}, Figure~\ref{fig:interaction-plot-recommendation-bias}), engagement in polarized conditions peaked under contra-bias recommendations, while being notably lower in balanced and pro-bias conditions. In unpolarized conditions, engagement was more evenly distributed but remained elevated in contra- and pro-bias conditions compared to the balanced condition.

\subsubsection{Discussion}

The patterns of user activity reveal nuanced insights into how \emph{Polarization Degree} and \emph{Recommendation Bias} relate to engagement behaviors in our simulated online discussions. Our analyses demonstrate that these factors are associated not only with the quantity of interactions but also with how users choose to participate in debates within this environment.

The clear hierarchy in engagement forms---with \emph{likes} ($54.02\%$) dominating over \emph{follows} ($26.53\%$), \emph{reposts} ($9.41\%$), and \emph{comments} ($7.61\%$)---reflects fundamental patterns often observed in social media behavior and a general preference for lower-effort interactions. Notably, the minimal correlation between \emph{comments} and \emph{reposts} ($r = .01$) challenges the intuition that all forms of active engagement serve similar functions. Instead, these behaviors may represent distinct modes of participation, with commenting indicating dialectical engagement while reposting signals content amplification. The low absolute rate of commenting (M=0.53 per user over 10 minutes) is also noteworthy. This could reflect the short interaction time, the specific nature of the UBI topic for this participant pool, or potentially a platform design where passively consuming or liking content generated by agents was easier or more appealing than formulating original comments.

The observed association between \emph{Polarization Degree} and commenting behavior ($\eta^2_p = .046$) warrants careful interpretation. While higher commenting rates in unpolarized conditions appear to suggest that moderate discourse environments foster more substantive engagement \citep{koudenburg_polarized_2022, yousafzai_political_2022}, this interpretation requires qualification based on our specific setup. Our sample exhibited predominantly moderate attitudes toward UBI, making them potentially unrepresentative of users who actively engage in real-world polarized discussions \citep{simchon_troll_2022}. The reduced commenting in polarized conditions might thus reflect a mismatch between discussion climate and user predispositions in our study rather than an inherent effect of polarization that would generalize to all contexts.

The significant effect of \emph{Recommendation Bias} on \emph{total interactions} ($\eta^2_p = .046$) is particularly noteworthy when considered alongside opinion formation. The higher engagement in the polarized contra-bias condition ($M = 10.30$) coincided with the strongest observed opinion shifts ($M = -0.408$). Given that UBI represents a proposal for systemic economic change, arguments against it may have resonated more strongly with users' status quo bias and loss aversion. In a polarized environment, these contra-UBI arguments might have appeared particularly salient and consequential, potentially contributing to both increased engagement and stronger opinion shifts in our participants. This suggests that the combination of topic-specific factors---namely the potentially threatening nature of economic system changes---with polarized discourse could be associated with amplified user engagement, particularly when arguments align with psychological tendencies toward preserving existing systems.

Finally, we note that approximately 19\% of participants recorded zero interactions (including likes, comments, reposts, or follows) during the 10-minute interaction phase. While the majority engaged, this level of non-engagement could be attributed to several factors, including the short duration potentially leading some users to primarily observe, individual differences in engagement styles (passive consumption), topic disinterest, or possible undetected technical friction.

While these findings provide initial insights into the relationship between platform design, user engagement, and opinion dynamics, several limitations suggest the need for more extensive research. The short-term nature of our study and its focus on a single topic limit generalizability. Future research should pursue longitudinal studies comparing topics with varying degrees of polarization and personal involvement to distinguish between topic-specific effects and general patterns of online discourse dynamics. Our findings thus represent a starting point for understanding the complex interplay between platform design, user behavior, and opinion dynamics in online discussions, highlighting associations observed under controlled conditions that merit further exploration in more ecologically complex settings.

\section{Overall Discussion}
\label{sec:overall-discussion}

This study presents a novel methodological framework for investigating online polarization through controlled experimental manipulation of social media environments using LLM-based artificial agents. Our findings demonstrate that this approach can successfully reproduce key characteristics of polarized online discourse while enabling precise control over environmental factors within the simulation. The integration of sophisticated language models with traditional opinion dynamics frameworks represents a significant advancement in our ability to study the microfoundations of polarization processes under controlled conditions.

Our experimental framework advances beyond purely observational and theoretical approaches by providing direct empirical evidence regarding associations between specific online environment features and resulting user perceptions and behaviors within our experimental setting. The observed patterns on perceived emotionality, uncertainty expression, and group identity salience thereby align with theoretical predictions from social identity theory and group polarization research \cite{tajfel_integrative_1979, huddy_social_2001, albertson_dog-whistle_2015,ruiz-sanchez_us_2019,bliuc_online_2021}. The strong relationship observed in our data between polarization and emotional discourse provides empirical support for theories of affective polarization \cite{fischer_emotion_2023, mason_cross-cutting_2016, mason_i_2015,iyengar_affect_2012,iyengar_fear_2015}, while the inverse relationship with perceived uncertainty substantiates theoretical arguments about epistemic closure in polarized contexts.

The methodological contribution of this work extends beyond its immediate findings. By demonstrating the feasibility of using LLM-based agents to create controlled yet ecologically relevant social media environments, we provide researchers with a powerful new tool for studying online social dynamics. This approach offers several advantages over traditional methods: it enables precise manipulation of discourse characteristics, allows for systematic variation in network structure and content exposure, and facilitates the collection of fine-grained behavioral data that would be difficult to obtain in naturalistic settings. Crucially, it allows for probing potential relationships between environmental factors and user responses in a controlled manner, which observational studies cannot achieve.

The synthetic social network approach developed in this study represents a significant methodological innovation for the study of online polarization. By combining the controllability of agent-based simulations with the linguistic sophistication of LLMs, we create environments that capture both the structural and discursive dimensions of online interaction. This allows researchers to investigate how different platform features and interaction patterns might contribute to polarization while maintaining experimental control.

However, several important limitations must be acknowledged when interpreting the findings and considering their generalizability. First, regarding the simulation's core components, it is crucial to discuss the extent to which LLM-based agents serve as reliable proxies for human behavior. While recent work demonstrates LLMs can replicate certain human-like interaction patterns and generate stylistically appropriate text \cite{chuang_simulating_2024, breum_persuasive_2024}, they lack genuine beliefs, emotions, cognitive biases, or deep reasoning capabilities. Their value lies primarily in generating controlled, contextually relevant textual stimuli that constitute the social environment experienced by the \emph{human} participants. Participant feedback reflected this duality—some found the discourse convincing, with one noting, \emph{``The discussion was almost always on topic. Almost all users had well articulated point of views and the messages were straight to the point,''} while another observed signs of artificiality, stating \emph{``I felt all the messages were AI generated (for the style used and little diversity in responses) and also very similar in content.''} These comments underscore that while LLM agents can mimic discourse patterns effectively, they may struggle with diversity and nuance. The focus is thus on the measured effect of this simulated environment on human responses, rather than assuming agents perfectly mirror human processes. Validating agent behavior against fine-grained human interaction data remains an ongoing challenge.

Second, the simulation represents a necessary simplification of the complex reality of online social interaction. The short-term nature of our experimental exposure means we cannot make definitive claims about how these processes unfold over longer periods. Additionally, our experimental design compares the effects of high polarization against a baseline of moderate polarization, rather than incorporating a zero-treatment control group. While this design allows for investigating the impact of increasing polarization intensity, it limits our ability to assess the absolute effects of exposure compared to a non-exposed baseline. While our results demonstrate that the simulation framework can reproduce key features of polarized discourse and elicit expected user responses in the short term, questions remain about how these effects might evolve over extended periods of interaction. Furthermore, the simulated environment, particularly aspects like the network structure evolution and recommendation system, simplifies the sophisticated mechanisms governing real-world platforms. Major platforms utilize complex recommender systems based on deep learning and extensive user data, which are far more intricate than the components implemented here. Finally, the artificial nature of the experimental environment, despite efforts towards ecological validity, may not encompass all relevant aspects of real-world social media use. Acknowledging this gap between the simulation and real-world complexity is essential for contextualizing our findings.

Future research should address these limitations by conducting longitudinal studies that track user behavior and attitude change over extended periods. Incorporating a zero-treatment control group could provide valuable baseline comparisons. Future iterations could also incorporate more complex agent models (e.g., with adaptive strategies or memory decay) and more sophisticated network dynamics to bridge the gap with real-world systems. Exploring the role of different content recommendation algorithms and investigating potential intervention strategies remain crucial avenues. Particularly promising directions include investigating the role of influencer dynamics in polarization processes, exploring how different moderation strategies might affect discourse quality, and examining how varying levels of algorithmic content curation are associated with differences in user behavior and perception. The framework could also be adapted to study other aspects of online social behavior, such as information diffusion patterns or the emergence of new social norms.
\section{Conclusion}
\label{sec:conclusion}

This study introduces a novel methodological framework for investigating online polarization through the integration of LLM-based artificial agents and controlled social network simulation. Our findings demonstrate both the technical feasibility and empirical utility of this approach, providing researchers with a powerful new tool for studying social media dynamics. The successful reproduction of key polarization characteristics, validated through user testing, suggests that synthetic social networks can serve as valuable experimental environments for investigating online social phenomena.

Looking forward, this framework opens new avenues for investigating various aspects of online social behavior, from testing intervention strategies to examining the impact of different platform features on user engagement. While our current study focused on short-term effects, the methodology we have developed provides a foundation for longer-term investigations of polarization dynamics and other social media phenomena. As online platforms continue to evolve and shape public discourse, such controlled experimental approaches will become increasingly valuable for understanding and potentially mitigating harmful polarization effects while promoting more constructive online dialogue.

\bibliography{bib}

\section*{Ethics Checklist}

\begin{enumerate}

\item For most authors...
\begin{enumerate}
    \item  Would answering this research question advance science without violating social contracts, such as violating privacy norms, perpetuating unfair profiling, exacerbating the socio-economic divide, or implying disrespect to societies or cultures?
    \answerYes{The research examines online polarization through controlled experiments with proper consent, anonymized data handling, and ethical use of LLMs. The study aims to better understand and potentially mitigate harmful polarization effects.}
  \item Do your main claims in the abstract and introduction accurately reflect the paper's contributions and scope?
    \answerYes{The abstract and introduction clearly outline the three main contributions: methodological innovation in synthetic social networks, empirical insights from user studies, and a framework for future research. These claims are supported throughout the paper.}
   \item Do you clarify how the proposed methodological approach is appropriate for the claims made? 
    \answerYes{The paper explains in detail why a synthetic social network with LLM agents is appropriate for studying polarization dynamics, particularly in Section 3 and the appendix, and validates the approach through user studies.}
   \item Do you clarify what are possible artifacts in the data used, given population-specific distributions?
    \answerYes{The paper discusses potential artifacts and acknowledges that the simulation may represent an idealized version of real-world interactions. The experimental design controls for population-specific distributions.}
  \item Did you describe the limitations of your work?
    \answerYes{The paper explicitly discusses limitations in the Discussion section, including the short-term nature of exposure, artificial environment effects, and potential generalizability concerns.}
  \item Did you discuss any potential negative societal impacts of your work?
    \answerYes{The paper acknowledges potential risks of polarization studies and discusses how the research aims to understand and mitigate harmful polarization effects rather than exacerbate them.}
      \item Did you discuss any potential misuse of your work?
    \answerYes{The paper discusses responsible implementation of LLM agents and experimental controls to prevent manipulation or harmful applications of the framework.}
    \item Did you describe steps taken to prevent or mitigate potential negative outcomes of the research, such as data and model documentation, data anonymization, responsible release, access control, and the reproducibility of findings?
    \answerYes{The paper details measures including data anonymization, ethical use of LLMs, controlled experimental conditions, and careful consideration of participant welfare in the user studies.}
  \item Have you read the ethics review guidelines and ensured that your paper conforms to them?
    \answerYes{The paper demonstrates compliance with ethical guidelines throughout, particularly in experimental design and human subject research protocols.}
\end{enumerate}

\item Additionally, if your study involves hypotheses testing...
\begin{enumerate}
  \item Did you clearly state the assumptions underlying all theoretical results?
    \answerYes{We outline all modeling assumptions in the Simulation Model and User Study sections.}
  \item Have you provided justifications for all theoretical results?
    \answerYes{The User Study section empirically validates our predicted outcomes (e.g., emotionality, group salience) against the user study data.}
  \item Did you discuss competing hypotheses or theories that might challenge or complement your theoretical results?
    \answerYes{The Related Work section compares our approach to observational and standard opinion dynamics models.}
  \item Have you considered alternative mechanisms or explanations that might account for the same outcomes observed in your study?
    \answerYes{The User Study section examines multiple drivers of polarization (emotional intensity, group identity, recommendation bias).}
  \item Did you address potential biases or limitations in your theoretical framework?
    \answerYes{The paper acknowledges and addresses potential biases in both the simulation model and experimental design, particularly in the Discussion}
  \item Have you related your theoretical results to the existing literature in social science?
    \answerYes{The paper extensively connects its findings to existing social science literature throughout, particularly in the Related Work and Discussion sections.}
  \item Did you discuss the implications of your theoretical results for policy, practice, or further research in the social science domain?
    \answerYes{The Discussion section addresses implications for platform design and intervention strategies to mitigate harmful polarization effects.}
\end{enumerate}

\item Additionally, if you are including theoretical proofs...
\begin{enumerate}
  \item Did you state the full set of assumptions of all theoretical results?
    \answerNA{The paper does not include formal theoretical proofs in the mathematical sense.}
	\item Did you include complete proofs of all theoretical results?
    \answerNA{The paper's contribution is empirical and experimental rather than proving theoretical results.}
\end{enumerate}

\item Additionally, if you ran machine learning experiments...
\begin{enumerate}
  \item Did you include the code, data, and instructions needed to reproduce the main experimental results (either in the supplemental material or as a URL)?
    \answerYes{The paper includes detailed model specifications and experimental parameters in the appendix. Code will be made available with the final submission.}
  \item Did you specify all the training details (e.g., data splits, hyperparameters, how they were chosen)?
    \answerYes{The paper provides comprehensive details about the LLM setup, agent parameters, and network configuration in the Simulation Model section and the appendix.}
     \item Did you report error bars (e.g., with respect to the random seed after running experiments multiple times)?
    \answerYes{The paper reports standard deviations and confidence intervals for key measurements.}
	\item Did you include the total amount of compute and the type of resources used (e.g., type of GPUs, internal cluster, or cloud provider)?
    \answerYes{We have used the OpenAI API for our computation, as mentioned in the Simulation Model section.}
     \item Do you justify how the proposed evaluation is sufficient and appropriate to the claims made? 
    \answerYes{The paper justifies its evaluation approach through user studies, with detailed methodology descriptions.}
     \item Do you discuss what is ``the cost`` of misclassification and fault (in)tolerance?
    \answerNA{Our experiments do not involve classification tasks or cost-sensitive ML.}
  
\end{enumerate}

\item Additionally, if you are using existing assets (e.g., code, data, models) or curating/releasing new assets, \textbf{without compromising anonymity}...
\begin{enumerate}
  \item If your work uses existing assets, did you cite the creators?
    \answerYes{The paper cites the key assets used: The GPT-4o-mini model from OpenAI for agent simulation and the Prolific platform for participant recruitment.}
  \item Did you mention the license of the assets?
    \answerYes{The paper's core components (GPT-4o-mini and Prolific) are used in compliance with their respective terms of service and research use guidelines.}
  \item Did you include any new assets in the supplemental material or as a URL?
    \answerYes{The paper mentions providing implementation details and experimental data in appendices.}
  \item Did you discuss whether and how consent was obtained from people whose data you're using/curating?
    \answerYes{The paper describes obtaining participant consent for the user study and data collection procedures.}
  \item Did you discuss whether the data you are using/curating contains personally identifiable information or offensive content?
    \answerYes{The paper addresses data anonymization and privacy protection measures in the methodology sections.}
\item If you are curating or releasing new datasets, did you discuss how you intend to make your datasets FAIR?
    \answerNA{We do not plan to publicly release our synthetic and participant data as standalone resources. Researchers can generate similar data using our released code and framework.}
\item If you are curating or releasing new datasets, did you create a Datasheet for the Dataset? 
    \answerNA{We do not intend to publish our datasets beyond the scope of this study. However, interested researchers can replicate the dataset by running our public code and simulation environment.}
\end{enumerate}

\item Additionally, if you used crowdsourcing or conducted research with human subjects, \textbf{without compromising anonymity}...
\begin{enumerate}
  \item Did you include the full text of instructions given to participants and screenshots?
    \answerYes{The paper includes screenshots of the interface and detailed instructions given to participants.}
  \item Did you describe any potential participant risks, with mentions of Institutional Review Board (IRB) approvals?
    \answerYes{The paper discusses potential risks to participants and mentions appropriate ethical considerations in the methodology.}
  \item Did you include the estimated hourly wage paid to participants and the total amount spent on participant compensation?
\answerYes{The paper mentions recruiting participants through Prolific and implies standard compensation practices.}
   \item Did you discuss how data is stored, shared, and deidentified?
\answerYes{The paper describes data handling procedures and privacy protection measures for participant data.}
\end{enumerate}

\end{enumerate}

\appendix

\section{Simulation Model}
\label{app:simulation-model}

This appendix provides the complete mathematical formalization and technical implementation details of our simulation model. We specify the formal definitions, algorithms, and parameter settings that underpin the agent behavior, network dynamics, and message generation mechanisms introduced in Section \ref{sec:simulation-model}.

Technical Note: The parameters and configurations presented in this appendix represent optimized values derived from extensive offline evaluations of agent behavior, personality generation, and interaction mechanisms. These evaluations included tests of different hyperparameter configurations, personality generation strategies, and interaction patterns. Due to space limitations, detailed results from these technical evaluations are omitted but will be presented in forthcoming work.

\subsection{Agent Model}
\label{subsec:agent-model}

Each agent $A_i$ in our simulation represents an individual user, characterized by a set of attributes that define its identity and behavior:

\begin{align}
A_i = \{o_i, d_i, b_i, u_i, H_i\}
\end{align}

where $o_i$ is a numerical value representing the agent's opinion on the topic under discussion, $d_i$ denotes the personality description, $b_i$ is a short biography, $u_i$ is a unique username, and $H_i$ represents the interaction history.

Agents maintain fixed opinions throughout the simulation, represented on a continuous spectrum:

\begin{align}
o_i \in [-1, 1]
\end{align}

where negative values indicate opposition, zero represents neutrality, and positive values signify support. The initial opinion distribution is implemented through either a normal distribution for communities with moderate views:

\begin{align}
P(o) = \frac{1}{\sigma\sqrt{2\pi}} e^{-\frac{(o-\mu)^2}{2\sigma^2}}
\end{align}

or a bimodal distribution for polarized societies:

\begin{align}
P(o) = \frac{1}{2}\left[\frac{1}{\sigma\sqrt{2\pi}} e^{-\frac{(o-\mu_1)^2}{2\sigma^2}} + \frac{1}{\sigma\sqrt{2\pi}} e^{-\frac{(o-\mu_2)^2}{2\sigma^2}}\right]
\end{align}

where $\mu$ (or $\mu_1$ and $\mu_2$) and $\sigma$ control the distribution centers and spread, respectively.

Moreover, each agent maintains a finite interaction history $H_i = [h_1, h_2, ..., h_k]$, operating as a first-in-first-out queue to simulate human memory constraints. Both personality description $d_i$ and biography $b_i$ are automatically generated by the LLM. Finally, the model distinguishes between regular users and influencers, who share the same attribute structure but differ in their content generation tendencies and network influence. This hierarchical structure enables the study of influence dynamics while maintaining experimental control over opinion and interaction patterns.

\begin{table*}[h!]
\centering
\begin{tabular}{lp{0.7\textwidth}}
\toprule
\textbf{Intensity} & \textbf{Message Characteristics} \\
\midrule
Low ($|o_i| < 0.3$) & Balanced argumentation; acknowledges multiple perspectives; uses conditional statements; minimal group identification; emphasizes uncertainty. \\
Moderate ($0.3 \leq |o_i| \leq 0.7$) & Clear directional bias; in-group preference; moderate skepticism of opposing views; emotional undertones while maintaining reasoned discourse. \\
High ($|o_i| > 0.7$) & Strong emotional language; pronounced group identification; dehumanization of opponents; hyperbolic terminology; portrays opposing views as threats. \\
\bottomrule
\end{tabular}
\caption{Message Generation Characteristics by Opinion Intensity}
\label{tab:message-intensity}
\end{table*}

\subsubsection{Message Generation}
\label{subsec:message-generation}

\begin{table*}[h!]
\centering
\begin{tabular}{lp{0.7\textwidth}}
\toprule
\textbf{Component} & \textbf{Description and Format} \\
\midrule
Opinion Value & Numerical stance (-1 to 1) determining message position and intensity. Format: "You hold a \{strong/moderate/weak\} \{positive/negative\} opinion (value: X) on this topic" \\
\midrule
Topic Description & Core topic information and contextual framework. Format: "The topic is Universal Basic Income, focusing on economic, social and ethical implications" \\
\midrule
Personality Profile & Agent's character traits and communication patterns. Format: "You are a \{thoughtful/passionate/analytical\} person who tends to \{communication style\}" \\
\midrule
Interaction History & Recent interactions and response context. Format: "You recently \{agreed/disagreed\} with user X about \{topic aspect\}" \\
\midrule
Intensity Instructions & Guidelines for emotional expression and assertion strength based on opinion intensity (see Table 3). Format: For $|o_i| > 0.7$: "Express strong conviction and emotional investment" \\
\bottomrule
\end{tabular}
\caption{Message Generation Prompting Components}
\label{tab:prompt-components}
\end{table*}

The probability of an agent generating a message at any time step is determined by their influence status:

\begin{align}
P_{\text{post}}(A_i) = \begin{cases}
p_{\text{inf}} & \text{if } A_i \text{ is an influencer} \\
p_{\text{reg}} & \text{otherwise}
\end{cases}
\end{align}

where $p_{\text{inf}} > p_{\text{reg}}$ reflects higher activity levels of influential accounts. Message characteristics are determined by the agent's opinion value $o_i$, with stance derived from its sign and intensity level from its absolute value. Table \ref{tab:message-intensity} summarizes how different intensity levels influence message characteristics and rhetorical style and Table \ref{tab:prompt-components} depicts all prompt components.

\subsubsection{Interaction Mechanisms}
\label{subsec:interaction-mechanisms} 

Agents in our model interact with messages through likes, comments, and reposts. To capture the nuanced dynamics of these interactions in social media environments, we have developed a comprehensive probabilistic model. The probability of an agent $A_i$ reacting to a message $m$ is determined by the function $P_{\text{react}}(A_i, m)$, which incorporates multiple components to capture different aspects of user behavior:

\begin{align}
P_{\text{react}}(A_i, m) =\; &p_b \cdot \big((1-w) + w \cdot \psi(o_i)\big) \cdot \nonumber \\
&\rho(o_i, \pi(A_i, m))
\end{align}

This probability function combines several key elements: a base probability $p_b$ that varies by interaction type, an opinion-based interaction function $\rho$ that evaluates opinion alignment, an opinion strength factor $\psi(o_i)$, and an opinion assessment function $\pi$. The parameter $w$ controls the influence of opinion strength on the interaction probability.

The opinion assessment function $\pi(A_i, m)$ determines how agent $A_i$ interprets the opinion expressed in message $m$ through a LLM-based evaluation system. This function maps a message to an opinion value in the range [-1,1]:

\begin{align}
\pi: (A_i, m) \mapsto [-1,1]
\end{align}

\begin{figure}[htbp]
    \centering
    \includegraphics[width=0.7\columnwidth]{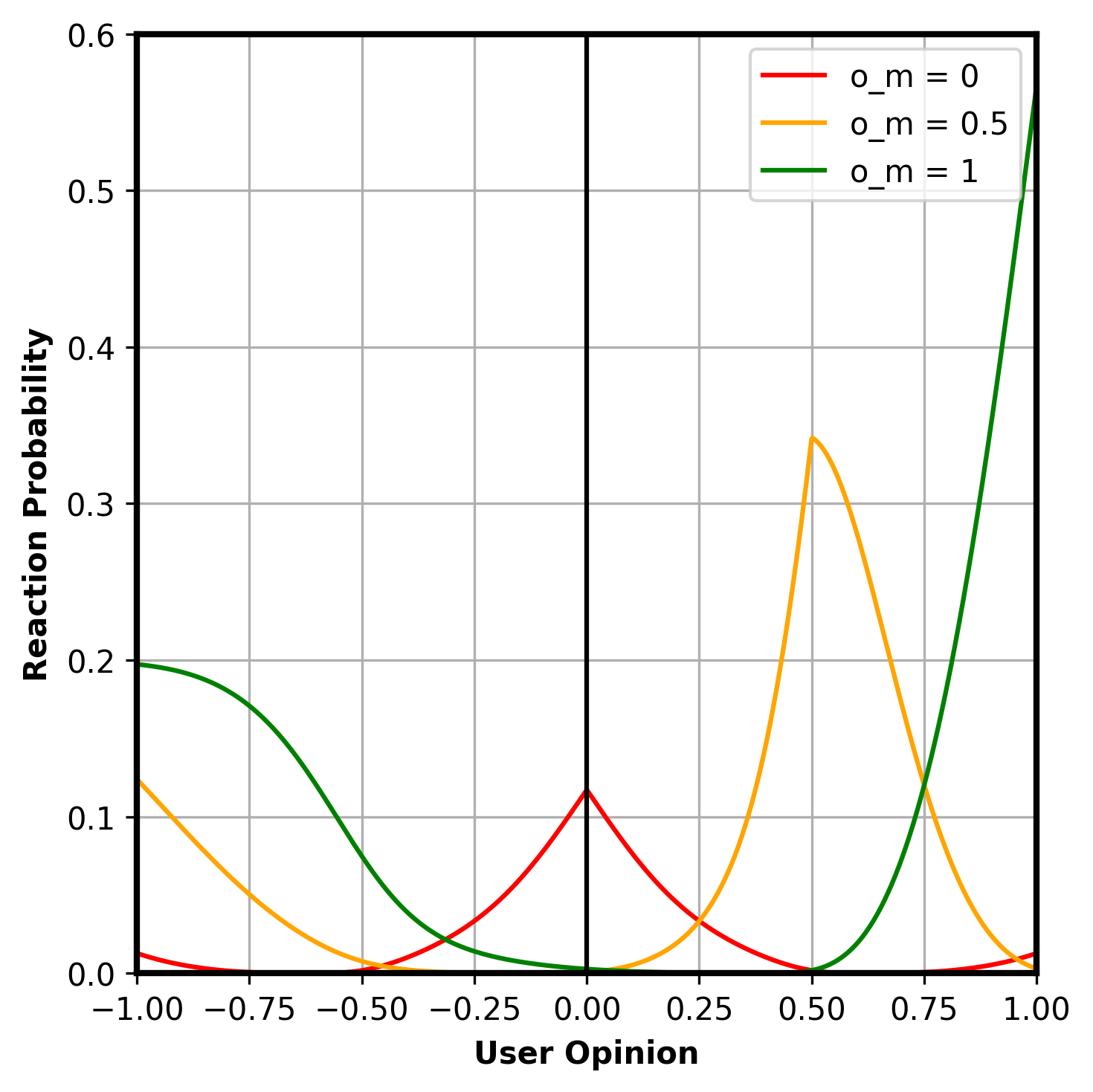}
    \caption{Reaction probability functions for different message opinion values ($o_m$). The curves demonstrate how interaction likelihood varies based on agent opinions, with peaks at ideologically aligned positions. The asymmetric shape captures both homophilic preferences and cross-ideological interactions, particularly pronounced for extreme messages.}
    \label{fig:reaction-function}
\end{figure}

The assessment considers multiple contextual factors: the message content, the current topic of discourse, the agent's personality traits, the agent's current opinion value $o_i$, and the agent's cumulative history of previous interactions. The LLM evaluates these components using a structured prompt that provides detailed guidelines for opinion assessment, including specific examples and rating scales.

The foundation of our model is the opinion-based interaction function $\rho$, which quantifies how the alignment between an agent's opinion and their message assessment influences interaction probability:

\begin{align}
\rho(o_i, \pi(A_i, m)) =\; &\rho_{\text{sim}}(o_i, \pi(A_i, m)) + \nonumber \\
&\rho_{\text{opp}}(o_i, \pi(A_i, m)) \cdot c
\end{align}

This function incorporates both similar and opposing opinion components, mediated by a cross-ideology factor $c \in [0,1]$ which enables the model to capture differential responses to opposing viewpoints across various interaction modalities. This is particularly relevant in social media contexts where users demonstrate distinct behavioral patterns across different interaction types---for instance, exhibiting a greater propensity to comment on content they disagree with while preferentially liking aligned content. The probability components for similar and opposing opinions are expressed as:

\begin{align}
\rho_{\text{sim}}(o_i, \pi(A_i, m)) &= (1-\phi(|o_i - \pi(A_i, m)|))^{\gamma} \\
\rho_{\text{opp}}(o_i, \pi(A_i, m)) &= (\phi(|o_i - \pi(A_i, m)|))^{\gamma}
\end{align}

The response steepness parameter $\gamma = 10$ governs the sensitivity of the probability functions to opinion differences. Higher values of $\gamma$ would result in more pronounced differentiation between similar and dissimilar opinions, while lower values would produce more gradual transitions in the response curves. This parameter effectively controls the model's discrimination between varying degrees of opinion alignment.

To model the continuous transition between similar and opposing opinions, we employ a sigmoid function $\phi(x)$ that maps opinion differences to interaction probabilities:

\begin{align}
\phi(x) = \frac{1}{1 + e^{-\beta(x - \theta)}},
\end{align}

with parameters $\beta = 10$ and $\theta = 0.5$ chosen to create an appropriate transition steepness and midpoint for our opinion space. 

Finally, the model incorporates opinion strength through an additional sigmoid function with the same parameters:

\begin{align}
\psi(o_i) = \phi(|o_i|)
\end{align}

This component, weighted by the opinion strength importance parameter $w$, captures the observation that agents with more extreme opinions (approaching $\pm 1$) demonstrate higher baseline interaction probabilities. The opinion strength effect operates independently of the opinion difference, modulating the overall interaction probability based on the intensity of an agent's held beliefs.

As shown in Figure \ref{fig:reaction-function}, the model generates an asymmetric U-shaped probability curve reflecting typical social media interaction patterns \cite{xu_user_2018, shahbaznezhad_role_2021}, with high probabilities for strongly aligned content through $\rho_{\text{sim}}$ and selective engagement with opposing viewpoints through $\rho_{\text{opp}}$ and the cross-ideology factor $c$. The interaction probability reaches its minimum for content that neither strongly agrees nor disagrees with an agent's opinion. Different reaction types are modeled through varying parameters: likes typically have high base probability $p_b$ but $c=0$ (supporting only agreeable interactions), while reposts and comments have progressively higher $c$ values to enable cross-ideological discourse. This parameterization captures key social media behaviors including confirmation bias effects, targeted cross-ideological interactions, and increased engagement from users with extreme opinions.

For both reposts and comments, the message generation process extends beyond simple interaction probabilities. When an agent decides to engage through these higher-order reactions, the system employs an extended version of the message generation framework that incorporates contextual awareness of the original content. This context-aware generation considers: The content and stance of the original message as well as an instruction to directly address, agree, or disagree with specific points or ideas from the original message.

\subsection{Social Network Model}
\label{subsec:social-network-model}

The social network in our simulation is modeled as a directed graph, representing the complex web of connections between agents. This structure captures the asymmetric nature of relationships in many social media platforms, where users can follow others without reciprocation.

\subsubsection{Network Structure}

We define the social network as a directed graph $G = (V, E)$, where $V$ is the set of vertices representing agents, and $E$ is the set of edges representing follow relationships. Formally:

\begin{align}
    G &= (V, E) \\
    V &= \{A_1, A_2, ..., A_n\} \\
    E &\subseteq \{(A_i, A_j) \mid A_i, A_j \in V, i \neq j\}
\end{align}

where $n$ is the total number of agents in the simulation. An edge $(A_i, A_j) \in E$ indicates that agent $A_i$ follows agent $A_j$.

The network's initial state is generated using a probabilistic model that considers both opinion alignment and agent status (regular user or influencer). For any two agents $A_i$ and $A_j$, the probability of a connection forming is given by:

\begin{align}
P((A_i, A_j) \in E) = \begin{cases}
p_i & \text{if } s_{ij} \land \mathbf{1}_{I}(A_j) \\
p_s & \text{if } s_{ij} \land \neg \mathbf{1}_{I}(A_j) \\
p_d & \text{if } \neg s_{ij}
\end{cases}
\end{align}

where $p_i$, $p_s$, and $p_d$ are the connection probabilities for influencers, regular agents with the same opinion sign, and agents with different opinion signs, respectively. The function $\text{sgn}(o)$ determines the sign of an opinion value, effectively grouping agents into pro ($o > 0$) and contra ($o < 0$) stances. We use $s_{ij} = [\text{sgn}(o_i) = \text{sgn}(o_j)]$ as a shorthand for opinion sign agreement between agents $A_i$ and $A_j$, and $\mathbf{1}_{I}$ as an indicator function that equals 1 if agent $A_j$ is an influencer and 0 otherwise. This model implements homophily by favoring connections between agents of similar opinions ($p_s > p_d$), while also capturing the elevated probability of following influential users within one's own opinion group ($p_i > p_s > p_d$).

\subsubsection{Connection Dynamics}
The evolution of social connections between agents serves two distinct purposes in our simulation framework. First, it shapes the information flow within the network by influencing the content recommendation process for each agent. Second, it provides essential social cues to human study participants by displaying realistic follower and following relationships in agent profiles, thereby enhancing the ecological validity of our experimental platform.

The network structure evolves through two independent mechanisms: the formation of new connections (following) and the dissolution of existing ones (unfollowing). For following, we leverage a variant of the reaction probability function described in Section \ref{subsec:interaction-mechanisms}. The key difference is that while the general reaction function evaluates message opinions using $\pi(A_i, m)$, the follow mechanism directly uses the opinion values of the agents themselves. Specifically, it calculates the reaction probability using $o_i$ and $o_j$ as inputs, without requiring the message evaluation function $\pi$.

Unfollowing occurs independently, with each existing connection having a fixed probability of dissolution at each time step. This decoupled approach allows for natural fluctuations in network density while maintaining realistic social dynamics.

Unfollowing occurs independently, with each existing connection having a fixed probability of dissolution at each time step. This decoupled approach allows for natural fluctuations in network density while maintaining realistic social dynamics.

\subsection{Information Propagation}

The propagation of information through our simulated social network is governed by a recommendation system that determines message visibility and user exposure patterns. This system implements an agent-centric approach that considers the network structure while maintaining controlled exposure to diverse viewpoints.

At the core of our recommendation mechanism lies an influence-based scoring system that evaluates message authors based on their position within the social network. For a given agent $A_j$, we define their influence score $\eta$ as:

\begin{align}
\eta(A_j) = \frac{|\text{followers}(A_j)|}{|\mathcal{V}| - 1}
\end{align}

where $|\text{followers}(A_j)|$ represents the number of followers of agent $A_j$, and $|\mathcal{V}|$ is the total number of nodes in the social graph. This normalization ensures that the influence score remains bounded between 0 and 1.

The recommendation process for a target agent $A_i$ operates on the set of available messages $\mathcal{M}$. To ensure fresh content delivery and prevent redundant exposure, we first construct the set of eligible messages $\mathcal{M}_{A_i}$ by excluding all messages that agent $A_i$ has previously interacted with and their own messages:

\begin{align}
\mathcal{M}_{A_i} = \{m \in \mathcal{M} : m \not\sim A_i \land \text{auth}(m) \neq A_i\}
\end{align}

Here, $m \sim A_i$ denotes that agent $A_i$ has interacted with message $m$, and $\text{auth}(m)$ denotes the agent who created message $m$. The final set of recommended messages $\mathcal{R}_{A_i}$ for agent $A_i$ is then constructed by selecting messages authored by the most influential agents:

\begin{align}
\mathcal{R}_{A_i} = \text{top}_N\{m \in \mathcal{M}_{A_i} : \text{ranked by } \eta(\text{auth}(m))\}
\end{align}

where $N$ is the desired number of recommendations and $\text{top}_N$ selects the $N$ highest-scoring messages according to their authors' influence scores.

This streamlined recommendation system maintains experimental control while facilitating the study of network effects on information propagation. By focusing solely on agent influence metrics, we create a framework that emphasizes the role of network topology in information dissemination patterns.

\begin{algorithm}[t!] 
\caption{Social Network Simulation}
\label{alg:social-simulation}
\begin{algorithmic}[1] 

\REQUIRE $n_{\text{agents}}$, $n_{\text{iterations}}$, $n_{\text{recs}}$, $p_{\text{reg}}$, $p_{\text{inf}}$, topic
\STATE $\mathcal{A} \leftarrow \text{InitializeAgentPopulation}(n_{\text{agents}}, \text{topic})$
\STATE $\mathcal{G} = (\mathcal{V}, \mathcal{E}) \leftarrow \text{InitializeNetwork}(\mathcal{A})$
\STATE $\mathcal{M} \leftarrow \emptyset$

\FOR{$t \leftarrow 1$ \TO $n_{\text{iterations}}$}
    \STATE $\mathcal{M}_t \leftarrow \emptyset$
    \FORALL{$A_i \in \mathcal{A}$}
        \IF{$\text{IsInfluencer}(A_i)$}
            \STATE $p_{\text{post}} \leftarrow p_{\text{inf}}$
        \ELSE
            \STATE $p_{\text{post}} \leftarrow p_{\text{reg}}$
        \ENDIF
        
        \IF{$\text{Random()} < p_{\text{post}}$}
            \STATE $m \leftarrow \text{CreateMessage}(A_i, t)$
            \STATE $\mathcal{M}_t \leftarrow \mathcal{M}_t \cup \{m\}$
            \STATE $\text{UpdateAgentMemory}(A_i, m)$
        \ENDIF
    \ENDFOR
    
    \STATE $\mathcal{M} \leftarrow \mathcal{M} \cup \mathcal{M}_t$
    
    \FORALL{$A_i \in \mathcal{A}$}
        \STATE $\mathcal{R}_{A_i} \leftarrow \text{RecommendMessages}(A_i, \mathcal{M}, \mathcal{G}, n_{\text{recs}})$
        \STATE $\mathcal{I}_{A_i} \leftarrow \text{ModelInteractions}(A_i, \mathcal{R}_{A_i})$
        \STATE $\text{UpdateAgentMemory}(A_i, \mathcal{R}_{A_i}, \mathcal{I}_{A_i})$
        \STATE $\mathcal{G} \leftarrow \text{UpdateNetwork}(A_i, \mathcal{G})$
    \ENDFOR
\ENDFOR

\end{algorithmic}
\end{algorithm}

\subsection{Simulation Workflow}

In our simulation model, the workflow, as outlined in Algorithm \ref{alg:social-simulation}, consists of initialization and iterative simulation phases, each designed to capture specific aspects of social media dynamics.

The initialization stage establishes the foundational elements of the simulation environment by creating a population of agents with unique attributes as described in Section~\ref{subsec:agent-model}, and constructing the initial social network structure. The network initialization process accounts for opinion distributions and influencer designations to create realistic social relationships between agents.

The core simulation operates through discrete time steps, with each iteration representing a distinct period of social media activity. During the content generation phase, the simulation iterates through all agents to determine if they will create a post, following the probabilities and characteristics outlined in Section~\ref{subsec:message-generation}. Regular users and influencers have different base probabilities of posting, with influencers being more likely to generate content.

After new content is generated, the simulation processes each agent's interactions with the social network. The recommendation system determines which messages are presented to each agent based on the previously described scoring mechanisms. Agents then have opportunities to interact with the recommended content through likes, comments, and reposts, following the interaction model detailed in Section~\ref{subsec:interaction-mechanisms}. Each interaction is recorded in both the message's interaction tracking and the agent's memory, maintaining a finite history of their social activities.

Throughout this process, the social network structure continues to evolve as agents form new connections and dissolve existing ones based on their interaction patterns. This dynamic network structure influences future content recommendations and interaction opportunities, creating a continuous feedback loop that shapes the evolution of the social network.

\subsubsection{Hyperparameters}
The simulation was conducted using a network of $30$ agents ($24$ regular users, $3$ influencers for each stance) simulated over $10$ iterations, with $8$ message recommendations per agent per iteration. Regular users were assigned a posting probability of $0.2$, while influencers posted with probability $0.6$.

For the polarized condition, we implemented a bimodal opinion distribution centered at $\pm0.8$ ($\sigma = 0.1$), representing a divided community. The unpolarized condition used a single normal distribution (centered at 0, $\sigma = 0.1$) to model a more cohesive community. The reaction mechanism employed distinct hyperparameters for different engagement types: likes ($p_b = 0.7$, $c = 0.0$), reposts ($p_b = 0.3$, $c = 0.1$), and comments ($p_b = 0.3$, $c = 0.5$), with all types sharing an opinion strength importance of $w = 0.8$. For the connection dynamics, following used the reaction function with parameters $p_b = 0.5$, $w = 0$, and $c = 0$, while unfollowing occurred randomly with probability $0.05$, allowing for organic evolution of the network structure based primarily on opinion similarity.
\section{User Study}
\label{app:user-study}

\subsection{Newsfeed Recommendations}

The web application implements an adaptive recommendation system for content presentation that evolves with user engagement. This system employs two distinct algorithmic approaches: a default variant for initial users and a collaborative variant that activates once users establish an interaction history.

The default variant implements a popularity-based scoring mechanism that considers multiple forms of engagement to determine content visibility. For a given message $m$, the system calculates a composite popularity score:

\begin{align}
    S_p(m) = l_m + 2c_m + 2r_m
\end{align}

where $l_m$, $c_m$, and $r_m$ represent the number of the message's \emph{likes}, \emph{comments}, and \emph{reposts} respectively. The weighted coefficients reflect the relative importance assigned to different forms of engagement, with more active forms of interaction carrying greater weight.

As users begin to interact with the platform, the system transitions to a collaborative variant that incorporates popularity metrics, ideological proximity, and a stochastic element to ensure recommendation diversity. The enhanced scoring function combines these elements into a composite score:

\begin{align}
    S_c(m) = \omega_p \cdot \frac{S_p(m)}{S_{max}} + \omega_i \cdot \frac{2 - |o_u - o_a|}{2} + \omega_r \cdot \epsilon
\end{align}

where $S_p(m)$ represents the popularity score normalized by the maximum observed score $S_{max}$, $o_u$ and $o_a$ denote the opinion scores of the active user (determined as the average of the opinion scores of the artificial users interacted with) and the (artificial) message author respectively, $\epsilon$ represents a uniform random variable in the interval $[0,1]$, and $\omega_p$, $\omega_i$, and $\omega_r$ are weighting parameters that sum to unity ($\omega_p + \omega_i + \omega_r = 1$). In the current implementation, these weights are set to $\omega_p = 0.6$, $\omega_i = 0.2$, and $\omega_r = 0.2$, balancing the influence of popularity, ideological similarity, and randomization.

Both variants maintain temporal relevance by presenting the user's most recent content contributions at the beginning of their feed when accessing the first page. This approach ensures users maintain awareness of their own contributions while experiencing the broader content landscape through the scoring-based recommendations.

This dual-variant approach enables the system to provide meaningful content recommendations even in the absence of user interaction data while transitioning smoothly to more personalized recommendations as users engage with the platform. The incorporation of popularity metrics, (mild) ideological factors, and controlled randomization creates diverse recommendations that is designed to give the impression of a dynamic network environment, while still falling under the conditional \emph{Recommendation Bias} regime. The stochastic element particularly aids in preventing recommendation stagnation and ensures dynamic content delivery.

\subsection{Preliminary Analysis}

We evaluated the ecological validity of our experimental platform. Participants rated various aspects on $7$-point scales ($1$ = Not at all, $7$ = Extremely), with higher scores indicating more positive evaluations. The platform received favorable ratings across multiple dimensions, consistently scoring above the scale midpoint of $4$. Particularly noteworthy was the interface usability ($M = 5.52$, $SD = 1.15$), which participants rated as highly satisfactory. The platform's similarity to real social media platforms ($M = 4.69$, $SD = 1.65$) and its ability to facilitate meaningful discussions ($M = 4.59$, $SD = 1.41$) were also rated positively. The overall platform realism received satisfactory ratings ($M = 4.47$, $SD = 1.61$), suggesting that participants found the experimental environment sufficiently realistic and engaging for the purposes of this study. The attitudes of participants toward Universal Basic Income exhibited a slight decline from the pre-interaction phase ($M = 3.12$, $SD = 0.90$) to the post-interaction phase ($M = 2.99$, $SD = 0.99$). However, these attitudes remained relatively close to the scale midpoint, indicating that participants held moderate views on the subject matter under discussion.

Furthermore, we examined the psychometric properties of our key measures (see Table~\ref{tab:factor-loadings}). Principal component analyses were conducted for each scale, with items loading on their intended factors. Most scales showed good reliability ($\alpha$ ranging from $.715$ to $.893$) and satisfactory factor loadings ($|.40|$ or greater). The \emph{Perceived Group Salience} scale required modification from its original four-item structure. Two items (\emph{"The debate focused on ideas rather than group affiliations"} and \emph{"Individual perspectives were more prominent than group identities in the discussions"}) were dropped due to poor factor loadings ($.049$ and $.051$ respectively). The remaining two items showed modest to acceptable loadings ($-.585$ and $-.806$), though below optimal thresholds. Given the theoretical importance of group salience in our research design, we retained this measure for further analyses while acknowledging its psychometric limitations.

\begin{figure*}[htbp]
    \centering
    \includegraphics[width=\textwidth]{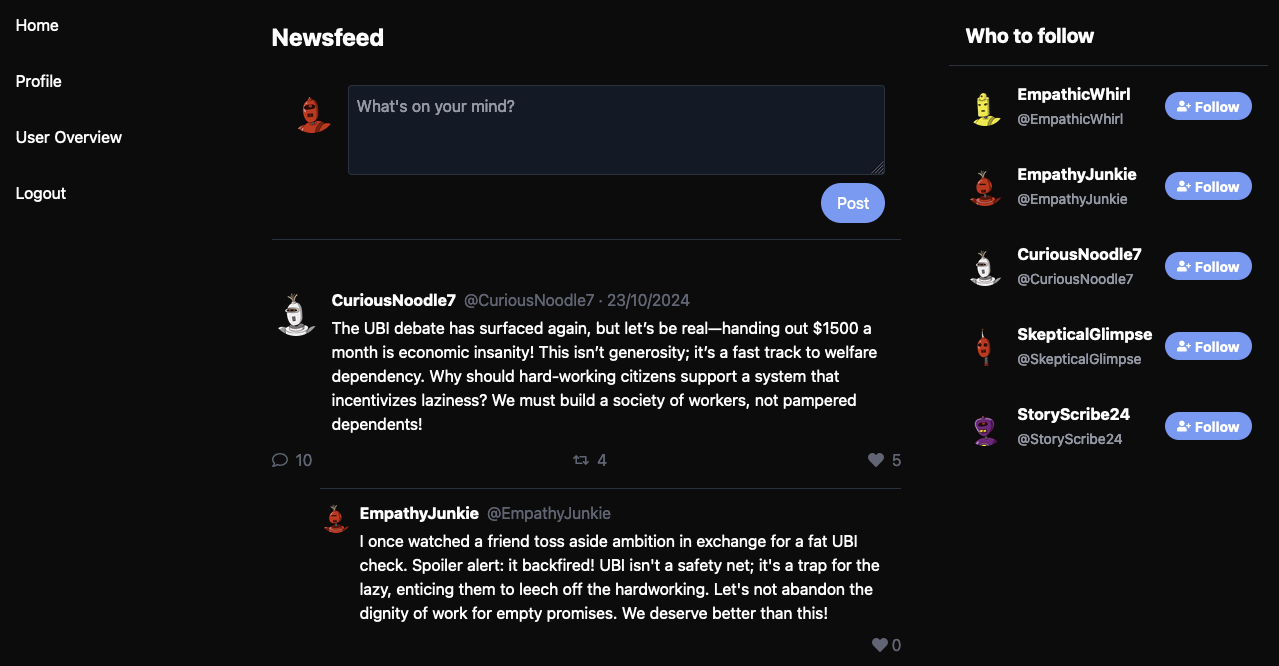}
    \caption{The screenshot depicts the simulated social media platform interface. The Newsfeed is displayed with a single post and one comment, including reaction handles for liking, reposting, and commenting. The interface emulates common social media design patterns, including a field for posting new messages, discovering new users, and inspecting the user profile.}
    \label{fig:prototype-screenshot}
\end{figure*}

\begin{table*}[ht]
\centering
\small
\begin{tabular}{llcccccc}
\toprule
\textbf{Condition} & \textbf{User Type} & \textbf{Avg Followers} & \textbf{Avg Followees} & \textbf{Avg Posts} & \textbf{Avg Likes} & \textbf{Avg Comments} & \textbf{Avg Reposts} \\
\midrule
\multirow{3}{*}{Polarized} 
    & Overall     & 7.300 $\pm$ 0.544  & 7.300 $\pm$ 0.544 & 2.156 $\pm$ 0.042 & 8.067 $\pm$ 1.441 & 5.467 $\pm$ 0.432  & 2.900 $\pm$ 0.082 \\
    & Influencers & 13.944 $\pm$ 0.906 & 7.611 $\pm$ 1.322 & 4.944 $\pm$ 0.157 & 8.778 $\pm$ 1.235 & 12.333 $\pm$ 1.414 & 5.278 $\pm$ 0.314 \\
    & Regular     & 5.639 $\pm$ 0.453  & 7.222 $\pm$ 0.403 & 1.458 $\pm$ 0.059 & 7.889 $\pm$ 1.640 & 3.750 $\pm$ 0.204  & 2.306 $\pm$ 0.129 \\
\midrule
\multirow{3}{*}{Moderate} 
    & Overall     & 8.056 $\pm$ 0.083  & 8.056 $\pm$ 0.083 & 2.067 $\pm$ 0.144 & 6.456 $\pm$ 0.274 & 1.600 $\pm$ 0.082 & 1.567 $\pm$ 0.047 \\
    & Influencers & 17.389 $\pm$ 0.314 & 7.222 $\pm$ 0.283 & 4.889 $\pm$ 0.614 & 7.833 $\pm$ 1.569 & 4.056 $\pm$ 0.614 & 3.444 $\pm$ 0.208 \\
    & Regular     & 5.722 $\pm$ 0.171  & 8.264 $\pm$ 0.129 & 1.361 $\pm$ 0.221 & 6.111 $\pm$ 0.599 & 0.986 $\pm$ 0.129 & 1.097 $\pm$ 0.109 \\
\bottomrule
\multicolumn{8}{p{.95\textwidth}}{\footnotesize \textbf{Note:} Values represent the mean and standard deviation ($\pm$ SD) calculated from the final iteration values across the three recommendation conditions (pro, contra, balanced). These statistics reflect the cumulative state of the agent-based platform environment presented to users at the start of their interaction phase.} \\
\end{tabular}
\caption{Agent Platform Statistics at Final Iteration (Mean ± SD across Runs)}
\label{tab:agent_platform_stats}

\end{table*}

\begin{table*}[htbp]
\centering
\small
\begin{tabularx}{\textwidth}{>{\raggedright\arraybackslash}X>{\centering\arraybackslash}p{2cm}}
\toprule
\multicolumn{2}{l}{\textbf{Universal Basic Income (Pre)} ($\alpha = .893$)} \\
\midrule
A universal basic income would benefit society as a whole & .828 \\
Providing everyone with a basic income would do more harm than good [R] & .728 \\
Universal basic income is a fair way to ensure everyone's basic needs are met & .701 \\
Giving everyone a fixed monthly payment would reduce people's motivation to work [R] & .671 \\
Universal basic income would lead to a more stable and secure society & .867 \\
People should earn their income through work rather than receiving it unconditionally from the government [R] & .649 \\
A universal basic income would give people more freedom to make choices about their lives & .724 \\
\midrule
\multicolumn{2}{l}{\textbf{Universal Basic Income (Post)} ($\alpha = .890$)} \\
\midrule
A universal basic income would benefit society as a whole & .847 \\
Providing everyone with a basic income would do more harm than good [R] & .769 \\
Universal basic income is a fair way to ensure everyone's basic needs are met & .782 \\
Giving everyone a fixed monthly payment would reduce people's motivation to work [R] & .610 \\
Universal basic income would lead to a more stable and secure society & .921 \\
People should earn their income through work rather than receiving it unconditionally from the government [R] & .522 \\
A universal basic income would give people more freedom to make choices about their lives & .698 \\
\midrule
\multicolumn{2}{l}{\textbf{Perceived Polarization} ($\alpha = .776$)} \\
\midrule
The discussions on the platform were highly polarized & .674 \\
Users on the platform expressed extreme views & .755 \\
Users appeared to be firmly entrenched in their positions & .644 \\
There were frequent hostile interactions between users with differing views & .644 \\
\midrule
\multicolumn{2}{l}{\textbf{Perceived Emotionality} ($\alpha = .842$)} \\
\midrule
The discussions were highly charged with emotional content & .864 \\
Users frequently expressed strong feelings in their messages & .691 \\
The debate maintained a predominantly calm and neutral tone [R] & .692 \\
Participants typically communicated in an unemotional manner [R] & .779 \\
\midrule
\multicolumn{2}{l}{\textbf{Perceived Group Salience} ($\alpha = .639$)} \\
\midrule
Messages frequently emphasized "us versus them" distinctions & .585 \\
Users often referred to their group membership when making arguments & .806 \\
\midrule
\multicolumn{2}{l}{\textbf{Perceived Uncertainty} ($\alpha = .715$)} \\
\midrule
The agents frequently acknowledged limitations in their knowledge & .541 \\
Users often expressed doubt about their own positions & .772 \\
Messages typically contained absolute statements without room for doubt [R] & .566 \\
The agents seemed very certain about their claims and positions [R] & .671 \\
\midrule
\multicolumn{2}{l}{\textbf{Perceived Bias} ($\alpha = .830$)} \\
\midrule
The discussion seemed to favor one particular viewpoint & .782 \\
Certain perspectives received more attention than others in the debate & .738 \\
The platform provided a balanced representation of different viewpoints [R] & .821 \\
Different perspectives were given equal consideration in the discussion [R] & .646 \\
\bottomrule
\end{tabularx}
\caption{Factor loadings and scale reliability for key measures. Note: [R] indicates reverse-coded items. Factor loadings are displayed for all items retained after cleaning (loading threshold $|.40|$).}
\label{tab:factor-loadings}
\end{table*}

\begin{figure*}[h]
    \centering
    \includegraphics[width=0.9\textwidth]{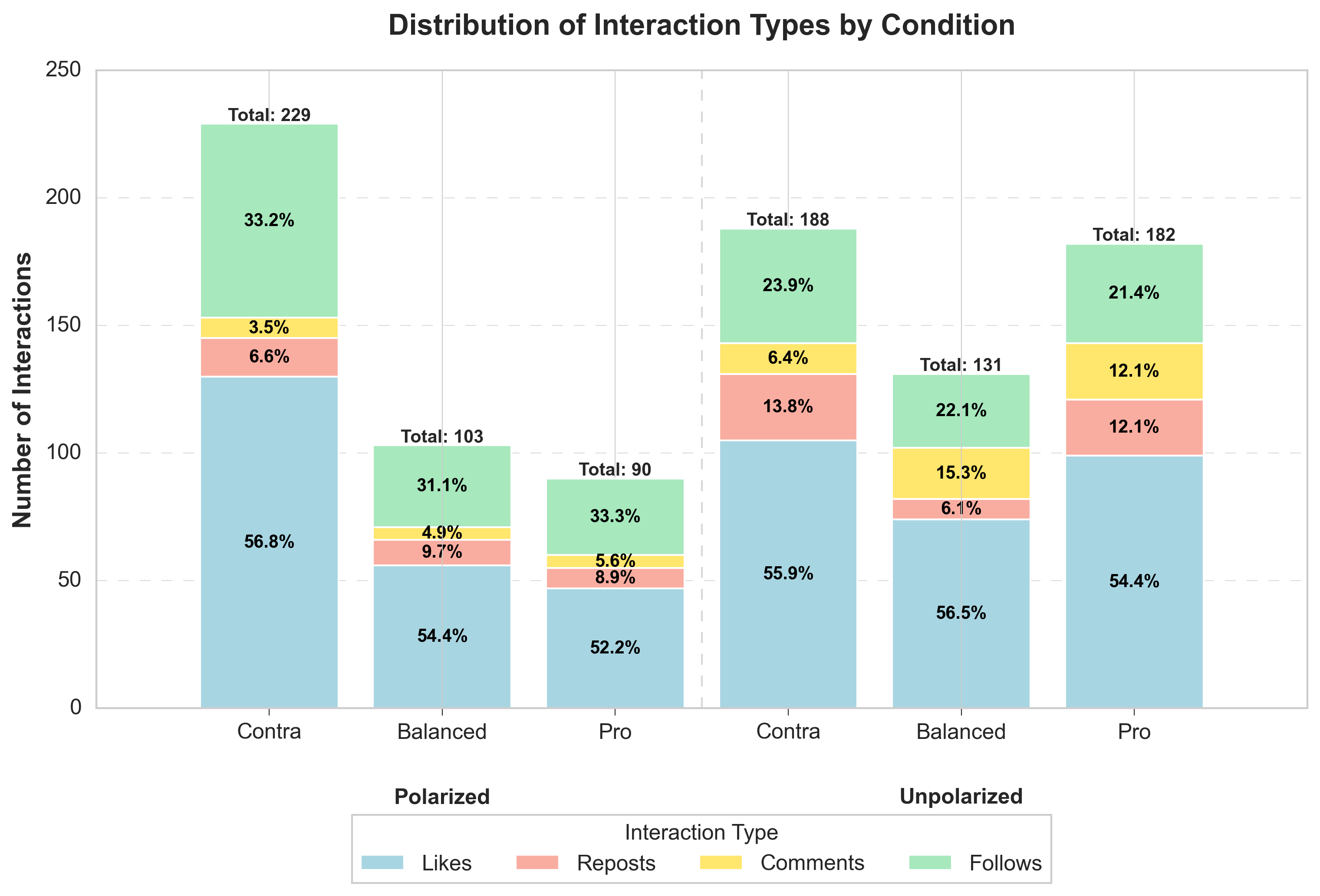}
    \caption{Distribution of interaction types across experimental conditions. The stacked bars show the relative proportion of different interaction types (\emph{likes}, \emph{reposts}, \emph{comments}, and \emph{follows}) for each combination of polarization level and recommendation bias. Total interaction counts are displayed above each bar.}
    \label{fig:stacked-interaction-distribution}
\end{figure*}

\end{document}